\documentclass[twocolumn,prl,showpacs,aps,superscriptaddress]{revtex4-2}

\usepackage{amsmath}
\usepackage{multirow}
\usepackage{graphicx} 
\usepackage{color}
\usepackage[latin1]{inputenc}
\usepackage{enumitem}

\begin{document}

\date{\today}

\title{Unraveling pairon excitations and the antiferromagnetic contributions \\in the cuprate specific heat}

\author{Yves Noat$^*$}

\affiliation{Institut des Nanosciences de Paris, CNRS, UMR 7588 \\
Sorbonne Universit\'{e}, Facult\'{e} des Sciences et Ing\'{e}nierie, 4 place
Jussieu, 75005 Paris, France}

\author{Alain Mauger}

\affiliation{Institut de Min\'{e}ralogie, de Physique des Mat\'{e}riaux et
de Cosmochimie, CNRS, UMR 7590, \\ Sorbonne Universit\'{e}, Facult\'{e} des
Sciences et Ing\'{e}nierie, 4 place Jussieu, 75005 Paris, France}

\author{William Sacks}

\affiliation{Institut de Min\'{e}ralogie, de Physique des Mat\'{e}riaux et
de Cosmochimie, CNRS, UMR 7590, \\ Sorbonne Universit\'{e}, Facult\'{e} des
Sciences et Ing\'{e}nierie, 4 place Jussieu, 75005 Paris, France}

\affiliation{Research Institute for Interdisciplinary Science,
Okayama University, Okayama 700-8530, Japan}



\begin{abstract}
Thermal measurements, such as the entropy and the specific heat,
reveal key elementary excitations for understanding the cuprates. In
this paper, we study the specific heat measurements on three
different compounds La$_{2-x}$Sr$_x$CuO$_4$,
Bi$_2$Sr$_2$CaCu$_2$O$_{8+\delta}$ and YBa$_2$Cu$_3$O$_{7-\delta}$
and show that the data are compatible with `pairons' and their
excitations. However, the precise fits require the contribution of
the antiferromagnetic entropy deduced from the magnetic
susceptibility $\chi(T)$.

Two temperature scales are involved in the excitations above the
critical temperature $T_c$: the pseudogap $T^*$, related to pairon
excitations, and the magnetic correlation temperature, $T_{max}$,
having very different dependencies on the carrier density ($p$). In
agreement with our previous analysis of $\chi(T)$, the $T_{max}(p)$
line is not the signature of a gap in the electronic density of
states, but is rather the temperature scale of strong local
antiferromagnetic correlations which dominate for low carrier
concentration. These progressively evolve into paramagnetic
fluctuations in the overdoped limit.

Our results are in striking contradiction with the model of J. L.
Tallon and J. G. Storey [Phys. Rev. B {\bf 107}, 054507 (2023)], who
reaffirm the idea of a $T$-independent gap $E_g$, whose temperature
scale $T_g=E_g/k_B$ decreases linearly with $p$ and vanishes at a
critical value $p_c \sim 0.19$.

Finally, we discuss the unconventional fluctuation regime above
$T_c$, which is associated with a mini-gap $\delta\sim$ 2\,meV in
the pairon excitation spectrum. This energy scale is fundamental to
the condensation mechanism.
\end{abstract}

\maketitle

\section{Introduction}

\begin{figure}[h!]
\vbox to 4 cm{
\includegraphics[width=8.0 cm, trim=10 10 10 10, clip]{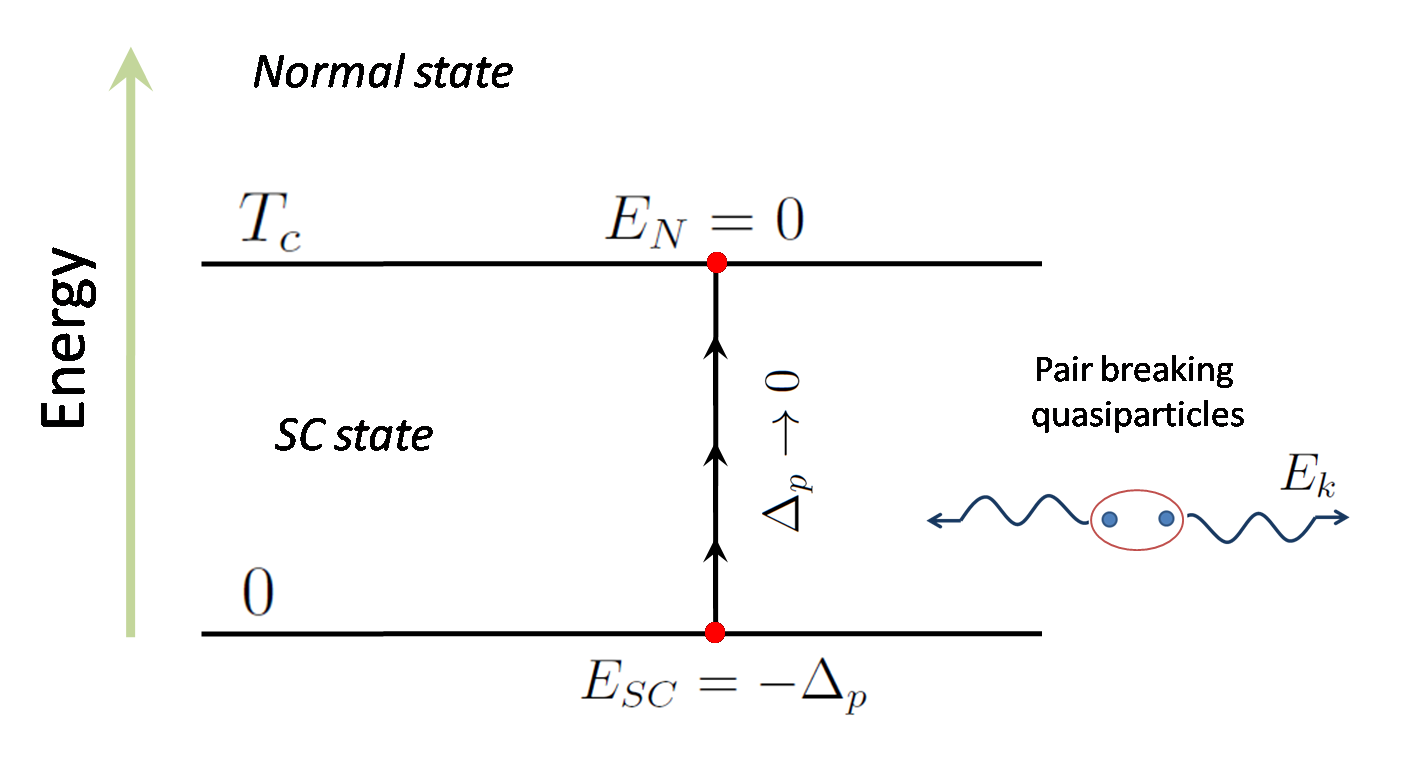}}
\caption{(Color online) Excitation process in a conventional
superconductor described by BCS theory \cite{PR_BCS1957}. Breaking a
Cooper pair gives rise to two quasiparticles with energy $E_k$. The
excitation is of fermionic type. As the temperature increases, more
and more quasiparticles are created (pair breaking) leading to the
decreasing of the gap $\Delta_p(T)$. The normal state is recovered
at the critical temperature $T_c$, where $\Delta_p(T_c)=0$. }
\label{Fig_BCS_scales}
\end{figure}

\begin{figure}[h!]
\includegraphics[width=5 cm, trim=10 10 10 10, clip]{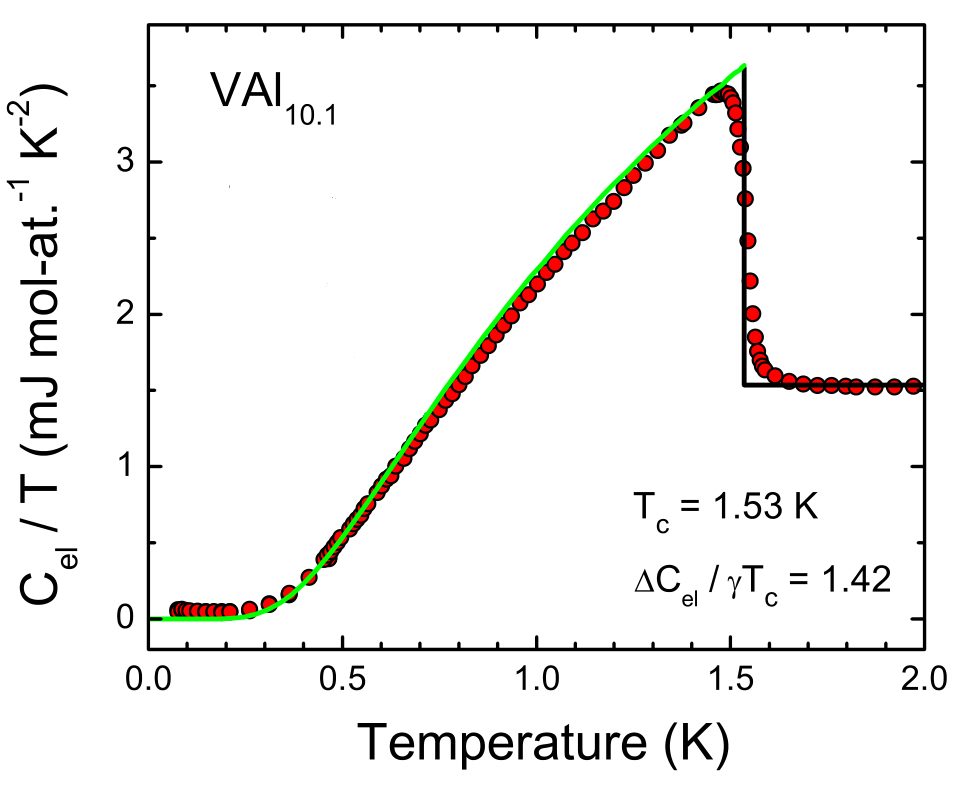}
\caption{(Color online) Experimental $\frac{C_v}{T}$ (red dots)
measured in VAl$_{10.1}$ by Klimczuk et al. (adapted from
Ref.\cite{JPhysCondMatt_Klimczuk2012}). Continuous line: Excellent
fit using BCS theory (see Ref. \cite{JPhysCondMatt_Klimczuk2012} for
details). } \label{Fig_Cv_Val}
\end{figure}

The measurement of the specific heat is a direct probe of thermal
excitations of a system. In the case of a conventional
superconductor, with negligible phonon contributions at low
temperature, the excitations are intimately linked to the nature of
the condensation (see Ref. \cite{ChiPhysB_Wen2020} for a review). In
the conventional case, described by the Bardeen-Cooper-Schrieffer
(BCS) theory \cite{PR_BCS1957}, elementary excitations are of the
fermionic type, the quasiparticles, as illustrated in
Fig.\,\ref{Fig_BCS_scales}. Moreover, the good agreement between
theory and experiment is striking, as in Fig.\,\ref{Fig_Cv_Val} from
Ref.\cite{JPhysCondMatt_Klimczuk2012}.


\begin{figure}
\vbox to 10.0 cm{
\includegraphics[width=7.0 cm, trim=10 10 10 10, clip]{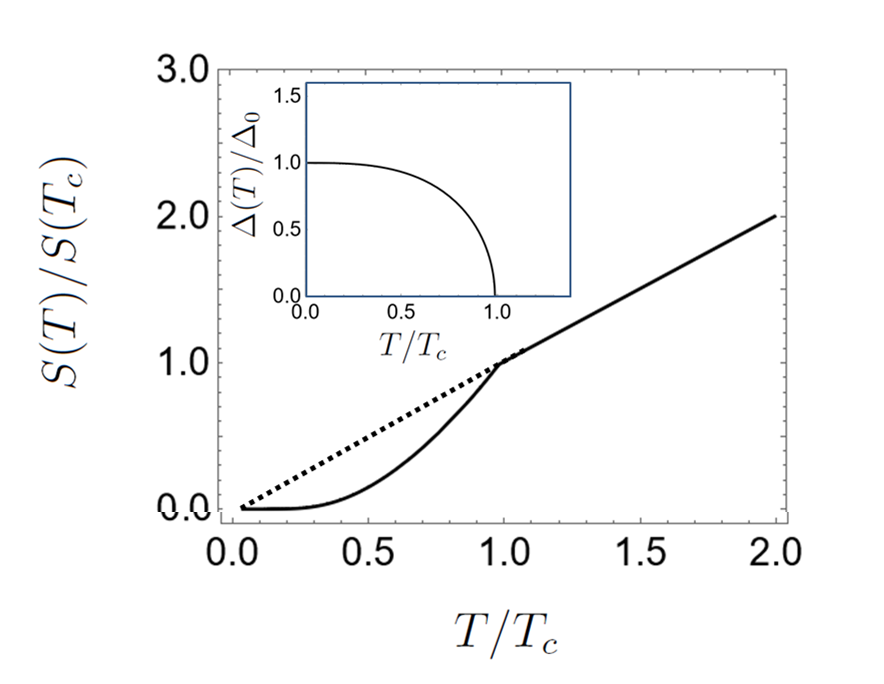}
\includegraphics[width=7.0 cm, trim=10 10 10 10, clip]{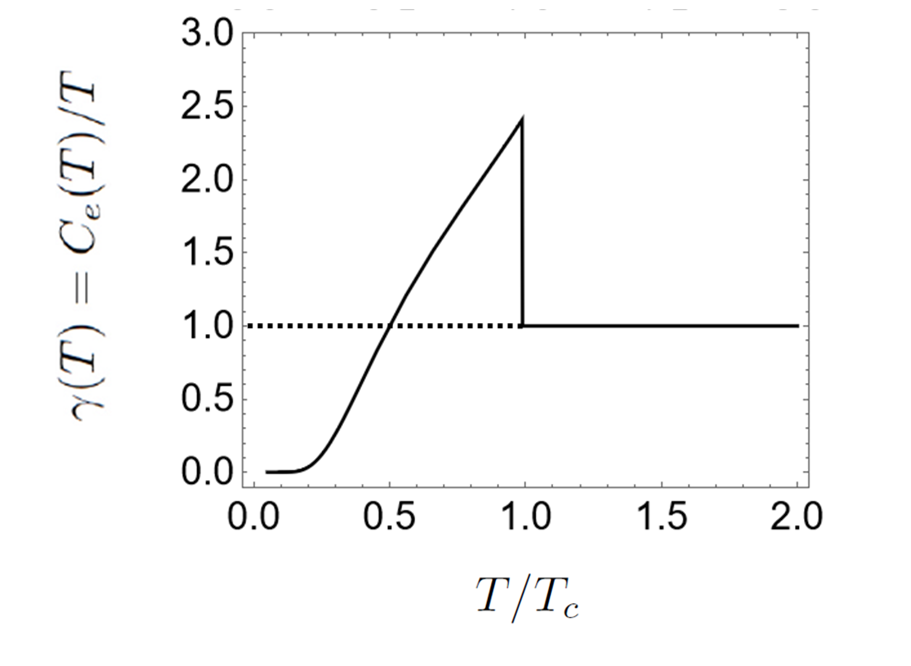}}
\caption{(Color online) Upper panel: Normalized entropy as a
function of temperature in the framework of the BCS theory
\cite{PR_BCS1957}. Inset: Normalized gap $\Delta(T)/\Delta_0$, where
$\Delta_0$ is the gap at $T=0$, as a function of temperature. Lower
panel: Coefficient $\gamma(T)=C_v(T)/T$ calculated in the framework
of the BCS theory. Note that the normal $\gamma_N$ is recovered
above $T_c$. } \label{Fig_Cv_BCS}
\end{figure}

The breaking of Cooper pairs gives rise to quasiparticles having the
dispersion relation $E_k=\sqrt{\epsilon_{k}^2+\Delta^2}$. The
occupation of quasiparticle states, given by the Fermi-Dirac
statistics, increases with temperature giving rise to a decrease of
the superconducting (SC) gap, which finally vanishes at $T_c$ (inset
of Fig.\,\ref{Fig_Cv_BCS}, upper panel). Above $T_c$ the normal
metallic state is fully recovered. It follows that there is a
discontinuity in the entropy slope and a corresponding jump in the
specific heat as shown in Figs.\,\ref{Fig_Cv_Val} and
\ref{Fig_Cv_BCS}. This is characteristic of a second order phase
transition and the vanishing of the order parameter, the energy gap,
at the critical temperature.

The situation is very different in cuprates where multiple degrees
of freedom are likely to be superposed (phonons, magnons, pair
excitations etc.) at the relevant temperatures. Several theoretical
works have addressed the difficult task to study the thermodynamic
properties, pointing out the important role of the pseudogap and
pair degrees of freedom above $T_c$
\cite{PRB_Moca2002,PRL_Curty2002,PRB_Borne2010}. Although these
contributions explore interesting concepts, so far they remain
inconclusive.

\begin{figure}
\includegraphics[width=8.0 cm, trim=10 10 10 10, clip]{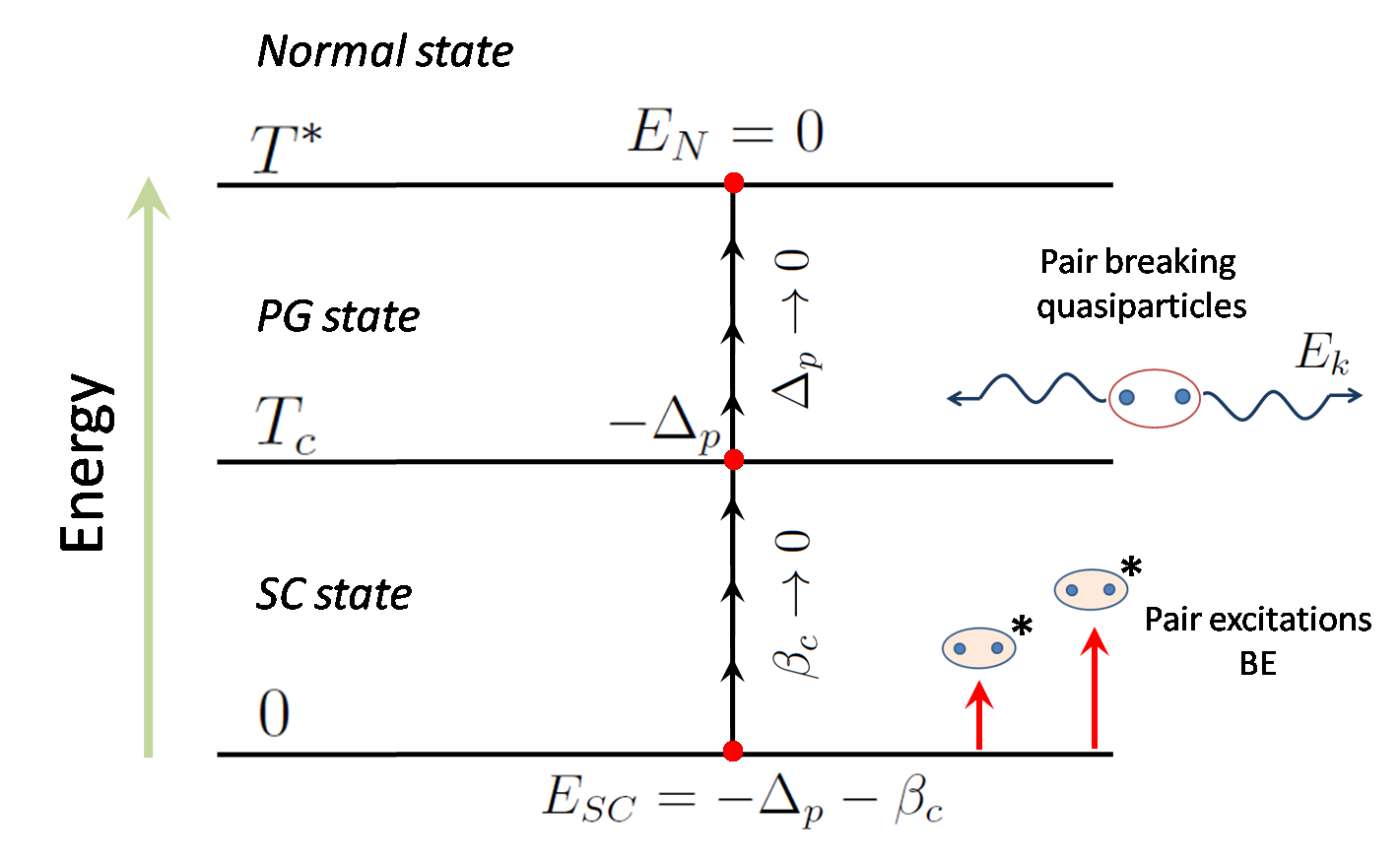}
\caption{(Color online) Upper panel: Excitation processes in a
cuprate superconductor in the framework of the pairon model
\cite{SolStatCom_Noat2021}. At low temperature, the dominant
excitations are pairon (bosonic) excitations. The diagram is
simplified in the underdoped case where pair excitations dominate
below  $T_c$ and quasiparticle excitations dominate above $T_c$. In
the general case, pair breaking quasiparticles can exist below
$T_c$. However, the critical temperature is still defined by the
vanishing of the pairon correlation energy $\beta_c$ and the
pseudogap temperature $T^*$ is defined by the vanishing of the
pairing gap. } \label{Fig_pairon_scales}
\end{figure}

\begin{figure}[h]
\vbox to 5.2 cm{
\includegraphics[width=7 cm, trim=10 10 10 10, clip]{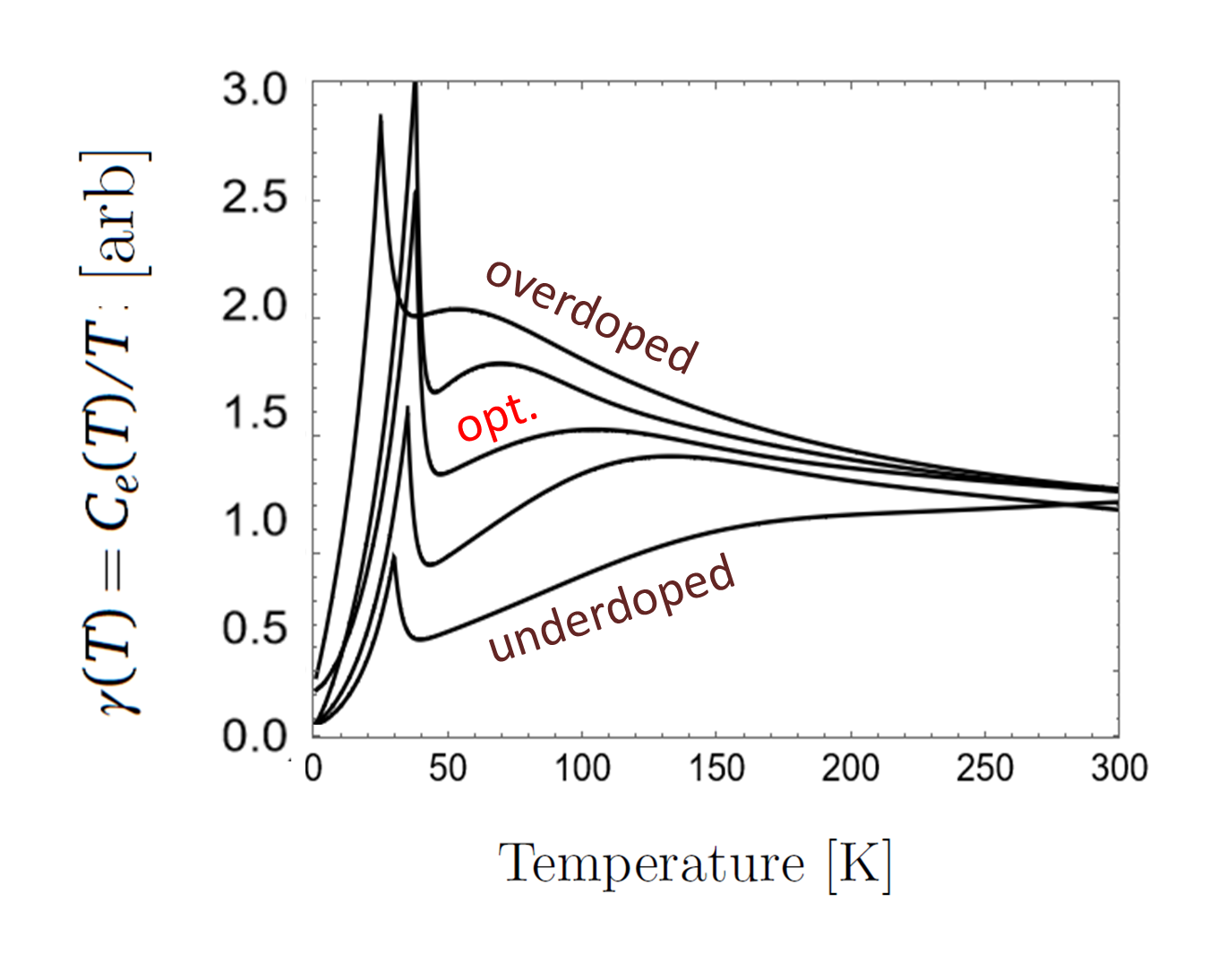}}
\caption{(Color online) General shape of $\gamma(T)$ coefficient for
different doping values from underdoped to the overdoped regime
using our phenomenological model described in the text. Notice the
continuous evolution of the background slope above $T_c$ and the
hump position.} \label{Fig_Cv_LSCO}
\end{figure}

In a previous work, we have calculated the contribution of pairon
excitations in the superconducting state as well as above $T_c$. The
relevant energy diagram is now shown in
Fig.\,\ref{Fig_pairon_scales}. The observed shape of the specific
heat strongly depends on the carrier concentration
\cite{PRL_Loram1993,PhysicaC_Loram1994,JphysJapSol_Matzusaki2004,PRL_Wen2009},
as illustrated in Fig.\,\ref{Fig_Cv_LSCO}, where the general
evolution of the temperature dependent $\gamma(T)$ coefficient is
shown for 5 different concentrations ranging from the underdoped to
the overdoped regimes. In the underdoped regime, above $T_c$,
$\gamma(T)$ is well below the normal state value up to very high
temperatures. This effect disappears in the overdoped regime, and
can even invert (upper curves in Fig.\,\ref{Fig_Cv_LSCO}). In
addition, in the whole doping range, there is no sharp discontinuity
in $C_v(T)$ at $T_c$, as expected in the BCS scenario (see Fig.
\ref{Fig_Cv_BCS}). Instead, an exponential tail is observed in
cuprates above $T_c$ on a typical temperature scale $\sim 5-10K$
\cite{PRB_Inderhees1987,PhysicaC_Loram1994,JphysJapSol_Matzusaki2004,PRL_Wen2009}
for LSCO and twice that for BSCO, which is one order of magnitude
above the fluctuations within the Ginzburg-Landau theory
\cite{PRB_Inderhees1987,PRB_Inderhees1988}.

In the underdoped regime, even at high temperatures, the normal
metallic state is not recovered. Instead, the entropy is roughly
linear at high temperature ($T\gtrsim 200K$), but below the expected
normal state line. This effect has been interpreted by Tallon et al.
in terms of a temperature-independent gap $E_g$ in the density of
states\,\cite{JPhysChemSol_Loram1998,JphysChem_Loram2001,FrontPhys_Tallon2022}.
In the specific heat, the latter is found to decrease with carrier
density and vanishes at a critical value $p_c=0.19$. However, this
is in direct contradiction with tunneling and photoemission
spectroscopic measurements (see
\cite{Revmod_Fisher2007,RepProgPhys_Hufner2008,Nat_Hashimoto2014,LowTemp_Kordyuk2015}
for reviews), where the pseudogap depends on temperature, vanishes
at $T^*$, and which does not cross the SC dome.

\begin{figure*}
\centering
\includegraphics[width=16 cm, trim=10 10 10 10, clip]{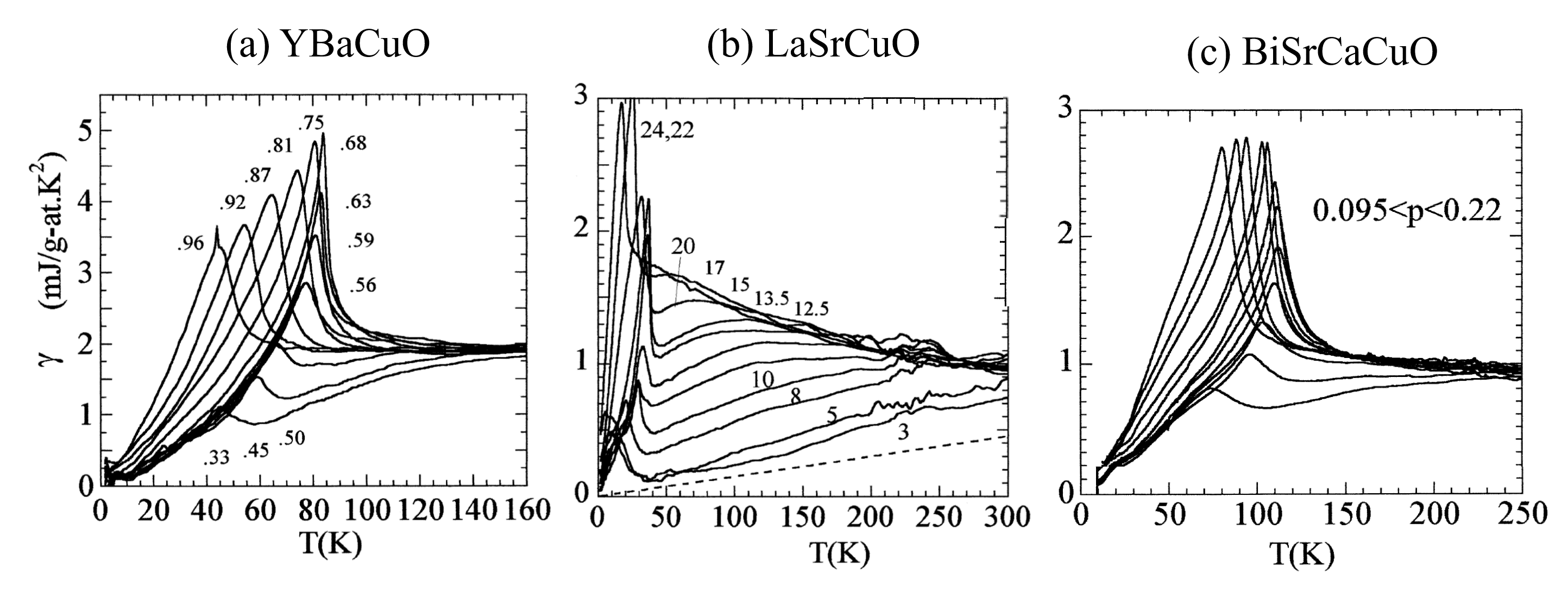}
\caption{(Color online) Experimental $\gamma(T)$ coefficient
measured in three different compounds \cite{JphysChem_Loram2001},
YBa$_2$Cu$_3$O$_{7-\delta}$ (panel a), La$_{2-x}$Sr$_x$CuO$_4$
(panel b), Bi$_2$Sr$_2$CaCu$_2$O$_{8+\delta}$ (panel c). }
\label{Fig_Cv_Exp}
\end{figure*}

In a previous article, we calculated the electronic contribution to
the specific heat in the framework of the pairon model
\cite{SolStatCom_Noat2021}. Two temperature scales are relevant, the
critical temperature $T_c$, as expected, and the pseudogap
temperature, $T^*$, at which the electronic state returns to the
normal metallic state. In the present article, we extend this work
to include two major effects: the magnetic excitations and the
fluctuations above $T_c$. This allows to do precise fits of the
specific heat of different materials La$_{2-x}$Sr$_x$CuO$_4$ (LSCO),
Bi$_2$Sr$_2$CaCu$_2$O$_{8+\delta}$ (BSCCO) and
YBa$_2$Cu$_3$O$_{7-\delta}$ (YBCO), see Fig.\,\ref{Fig_Cv_Exp}. In
particular, we show that in addition to $T_c$ and  $T^*$, there is a
third temperature scale $T_{max}\sim J/k_B$,  the characteristic
temperature of antiferromagnetic (AF) correlations, where $J$ is the
AF exchange energy, in agreement with our study of magnetic
susceptibility \cite{Solstatcom_Noat2022}, as well as pioneering
works \cite{PRB_Torrance1989,SolstatCom_Oda1990,PRB_Nakano1994}.

From this analysis, the behavior of the entropy in the underdoped
regime described above is due to the magnetic contribution and not
to a gap in the electronic excitations. This is in stark
contradiction with the conclusions advanced by Tallon and Storey
\cite{PRB_Tallon2023} who rule out the existence of excited pairons
in the specific heat.

\section{The pairon model}

The following important issues are still under debate in cuprates:
\begin{itemize}
\item[--] the pairing mechanism,
\item[--] the nature of the pseudogap and its relation to superconducting
order,
\item[--] the mechanism leading to the SC condensation,
\item[--] and the dependence of the physical parameters on carrier
concentration.
\end{itemize}

These issues can be addressed in the framework of our pairon model.
First, we have proposed that in cuprates, hole pairs, or pairons,
are formed below the characteristic temperature $T^*$ due to the
persistence of local magnetism on a typical scale $\xi_{AF}$, the
antiferromagnetic (AF) correlation length \cite{EPL_Sacks2017}.
Their binding energy is directly related to the effective AF
exchange energy $J_{eff}$ due to the local magnetic order
surrounding the pairon. At low temperature, pairons condense in a
collective quantum state as a result of their mutual interactions
\cite{SciTech_Sacks2015}.

There are two fundamental energy scales, the antinodal energy gap
$\Delta_p$, associated with pair formation, and the coherence energy
$\beta_c$, associated with pair correlations. These two energies are
proportional to the two temperature scales $T^*$, the pseudogap
temperature, and $T_c$, the critical temperature, respectively
\cite{PhysLettA_Noat2022}:
\begin{eqnarray}
&& \Delta_p=2.2\,k_B\,T^*
  \nonumber\\
&& \beta_c=2.2\,k_B\,T_c
\label{Equa_Ener_Temp}
\end{eqnarray}
The pseudogap temperature $T^*$ corresponds to the onset temperature
of pairon formation while the critical temperature $T_c$ corresponds
to the onset of pairon-pairon correlations.

In addition, in our view, the pseudogap and the superconducting
order are intimately linked \cite{PhysLettA_Noat2022}. Indeed, we
have shown that $T^*$ and $T_c$  can be expressed as follows
\cite{ModelSimul_Noat2022}, in terms of the hole concentration:

\begin{eqnarray}
&&T^*=\alpha_1 \,(1-p^\prime)  \nonumber\\
&& T_c=\alpha_2 \,p^\prime(1-p^\prime)
\label{Equa_Tstar_Tc}
\end{eqnarray}
where the reduced density is:
$$p^\prime=\frac{p-p_{min}}{p_{max}-p_{min}}$$ with $p_{min}=0.05$,
the minimum doping value and $p_{max}=0.27$, the end of the SC dome.
$\lambda_i$, $i=1,2$, are constants.

For BSCCO, the above relations are in very good agreement with
experiments. Within the experimental resolution, we found
$\alpha_1\simeq\alpha_2\approx 390$\,K. For LSCO $\alpha_1$ and
$\alpha_2$ are found to have slightly different values
\cite{ModelSimul_Noat2022} ($\alpha_1/\alpha_2=1.25$ with
$\alpha_1=200$\,K). Both $\alpha_1$ and $\alpha_2$ are proportional
to the same energy, $J_{eff}$.

At zero temperature, pairons form an ordered SC state. At finite
temperature, they are excited in higher pair energy states. In
addition, excited pairons can decay into quasiparticles. Thus, both
bosonic (excited pairs) and fermionic (quasiparticles) excitations
are present  \cite{SolStatCom_Noat2021}, as illustrated in
Fig.\,\ref{Fig_pairon_scales}. Contrary to the BCS case, the energy
gap is not the order parameter. Indeed, since pairons exist above
$T_c$, the critical temperature (and therefore the order parameter)
is determined by pairon-pairon correlations, leading to the $T_c$
dome \cite{PhysLettA_Noat2022}. Although the pair excitations follow
Bose-Einstein statistics, the  $T_c$ is not determined by the
standard expression for non-interacting bosons. Rather, the
condensate density, which is proportional to the correlation energy
$\beta_c(T)$ \cite{SciTech_Sacks2015}, is the true SC order
parameter.

So far this model fits very well the $T_c(p)$ and $T^*(p)$ phase
diagram, and fits quantitatively many experiments such as tunneling
\cite{SciTech_Sacks2015,EPJB_Sacks2016} ARPES
\cite{Jphys_Sacks2018,EPL_Noat2019} and magnetic susceptibility
\cite{Solstatcom_Noat2022}.

\newpage

\subsection{Entropy in the pairon model}


In order to understand the shape of the experimental curves, we now
recall the ingredients of the electronic entropy of pairons. The
details of the calculation can be found in our previous article
\cite{SolStatCom_Noat2021}. The entropy $S(T)$ results from the
contribution of both bosonic and fermionic excitations. At finite
temperature, pairons are excited out of the condensate, their
occupation being given by the Bose-Einstein statistics. $S(T)$ can
be written in a concise way:
\begin{equation}
S(T)=\sum_i n_i(\varepsilon_i,T) S_i(\varepsilon_i,T)
\label{Eq_entropy}
\end{equation}

where $n_i$ is the density of excited pairons with energy
$\varepsilon_i$ with the associated entropy term $S_i$. Since
pairons are composite bosons, we have  $n_i\propto
f_{BE}(\varepsilon_i,T)P_0(\varepsilon_i)$, where
$f_{BE}(\varepsilon)=1/\left(\exp\left(\frac{\varepsilon-\mu_b}{k_BT}\right)-1
\right)$ is the Bose-Einstein distribution and $P_0(\varepsilon_i)$
is the excited pair density of states.

\begin{figure}
\includegraphics[width=8. cm, trim=10 10 10 10, clip]{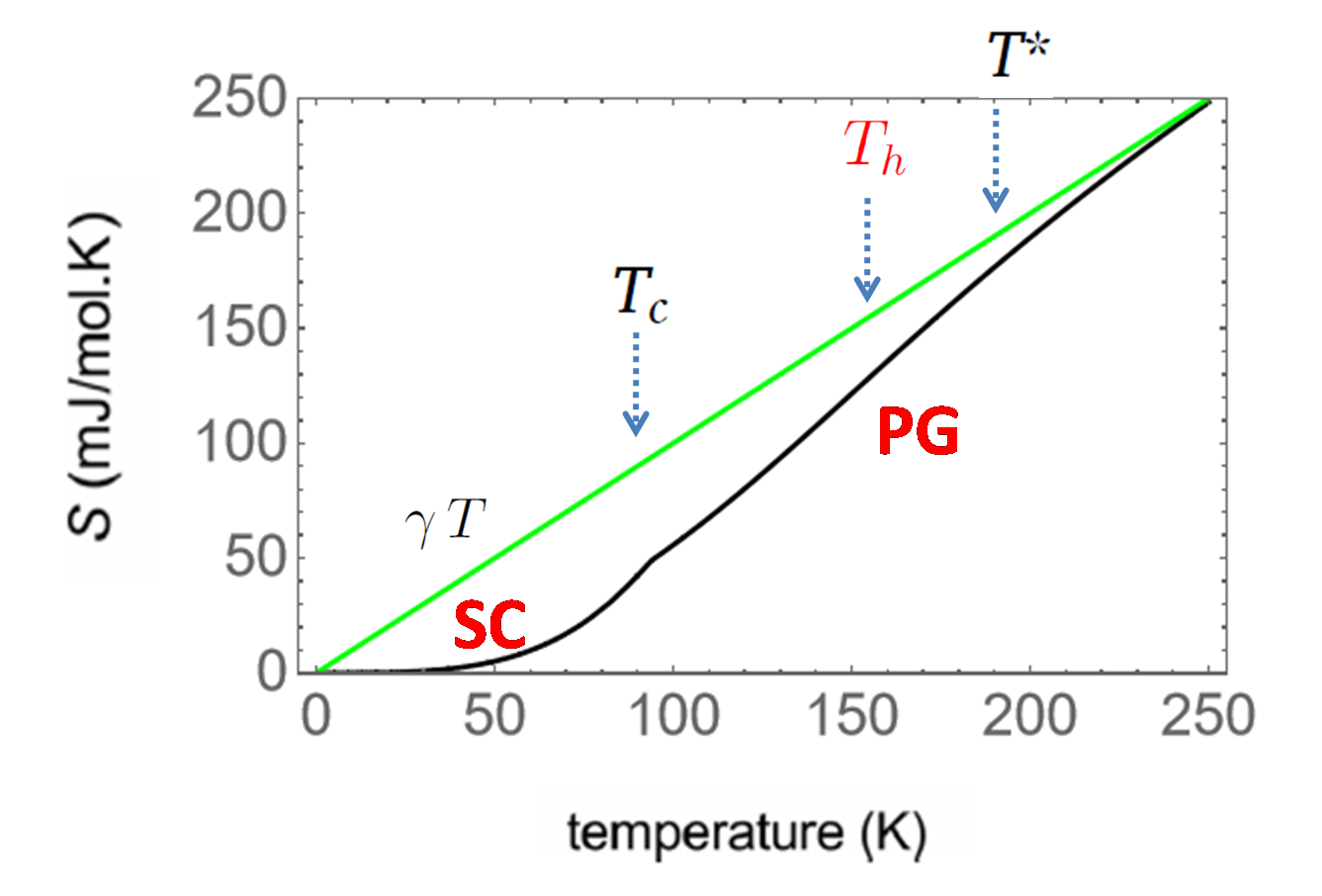}
\caption{(Color online) Plot of the entropy with temperature in the
pairon model revealing the important temperature scales (values for
BSCCO). The entropy has a discontinuous derivative at $T_c$ due to
the disappearance of the condensate but does not join the normal
state $\gamma_N T$ line. A weak inflection above $T_c$ is
characterized by the temperature $T_h$. The normal state entropy is
recovered at the pseudogap temperature $T^*$, indicating the
vanishing of the pairing gap. } \label{Fig_Pairon_Temp}
\end{figure}

Each excited pairon with energy $\varepsilon_i$ can decay into
quasiparticles with energy $E_k^i$.  Thus, $S_i$ can be expressed as
a sum over each quasiparticle contribution
\begin{equation}
S_i(\varepsilon_i,T)=\sum_{\vec{k}}S(E_k^i,T)
\label{Eq_Si}
\end{equation}
where $S(E_k^i,T)$ is the fermionic entropy associated with the
quasiparticle of energy
$E_k^i=\sqrt{\epsilon_{k}^2+{\Delta_k^i}^2}$. The excited pairs are
related to the pairing amplitudes via the equation: $\varepsilon_i =
\Delta_k^i- \Delta_p$ (see \cite{Solstatcom_Noat2022} and
\cite{SolStatCom_Noat2021}). As seen in Fig.\,\ref{Fig_Pairon_Temp},
this entropy takes into account not only the transition at $T_c$,
where the derivative of the entropy is discontinuous, but also the
return to the normal state through the end of the pseudogap at $T^*$
\cite{SolStatCom_Noat2021}.

\subsection{Fluctuations observed above $T_c$}

A remarkable and general feature of the $\gamma(T)$ experimental
curves is the prominent exponential decay just above the critical
transition. We can extend the above calculation of $S(T)$ in a
phenomenological way in order to take into account this fluctuation
regime above $T_c$.

\begin{figure}
\includegraphics[width=7.0 cm, trim=10 10 10 10, clip]{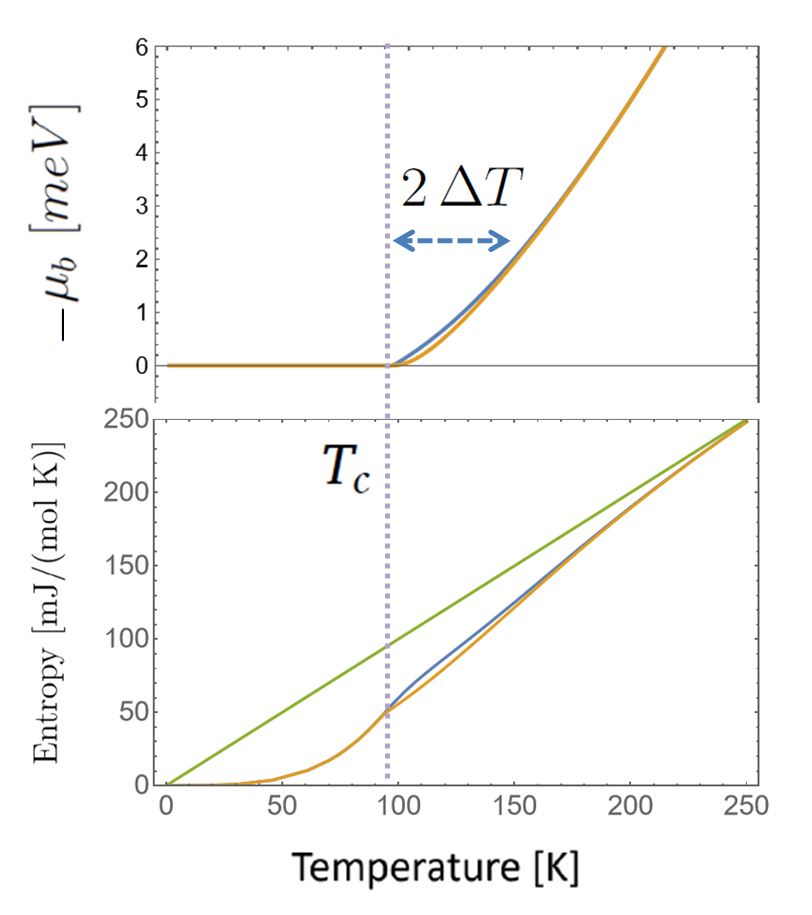}
\caption{(Color online) Upper panel: chemical potential (orange
line) and modified chemical potential (blue line) including
fluctuations in such a way that the entropy is smoothed above $T_c$
on the scale $\Delta T$ (values for BSCCO). Lower panel: the
corresponding entropy using the chemical potentials of the upper
panel. With the standard chemical potential (orange line), the
derivative of $S(T)$ is discontinuous at $T_c$ while, with the
modified chemical potential (blue line), it is continuous at $T_c$.
Green line: normal state entropy.} \label{Fig_ChemPot}
\end{figure}

\begin{figure}
\includegraphics[width=6.0 cm, trim=10 10 10 10, clip]{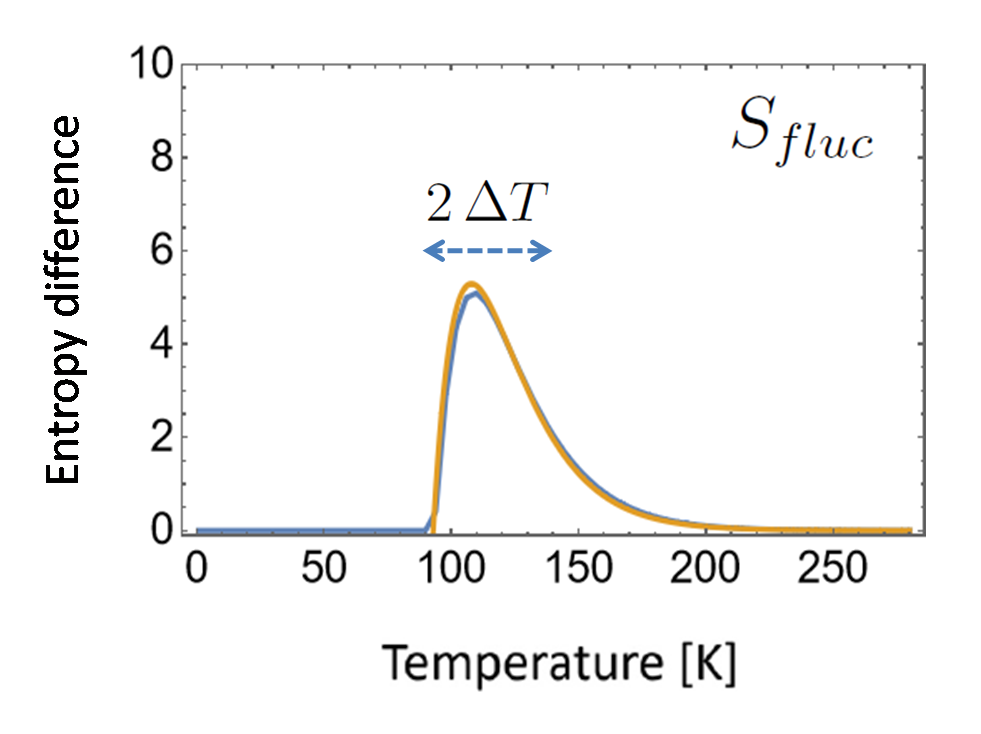}
\caption{(Color online) Fluctuation entropy $S_{fluct}$ (blue curve)
determined by subtracting the two entropy curves in Fig.
\ref{Fig_Cv_num}. $S_{fluct}$ sharply increases above $T_c$, reaches
a maximum and then decreases rapidly on the typical temperature
scale $\Delta T\sim 10-20K$ (values for BSCCO). Orange curve:
$S_{fluct}$ determined by the phenomenological expression, Eq.
\ref{Equa_Sfluc}.} \label{Fig_S fluc}
\end{figure}

\begin{figure}
\includegraphics[width=8.0 cm, trim=10 10 10 10, clip]{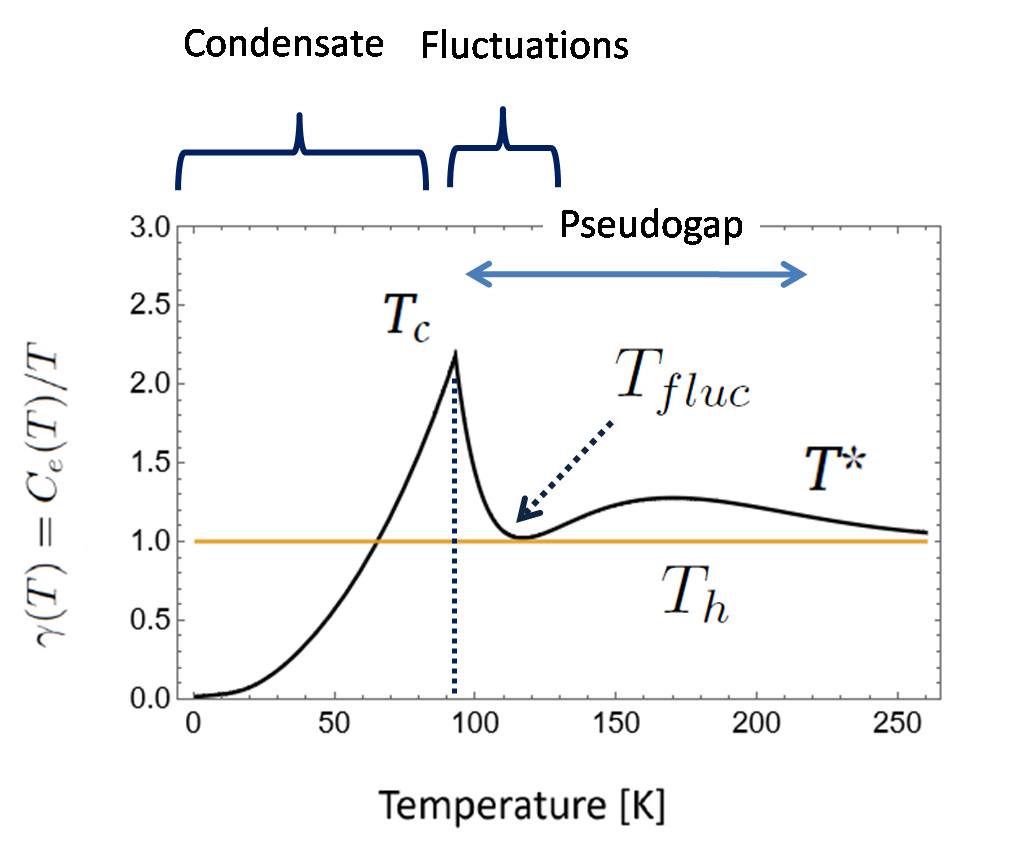}
\caption{(Color online) Electronic specific heat calculated in the
framework of the pairon model, including the fluctuations. Three
temperature scales can be distinguished above $T_c$: the fluctuation
temperature $T_{fluc}$, the temperature of the hump $T_h$ and the
pseudogap temperature $T^*$ where the normal state is recovered
(values for BSCCO).} \label{Fig_Cv_num}
\end{figure}

One approach, inspired by Ref. \cite{Adv_Kocharovsky2006}, is to
consider a slightly modified chemical potential $\mu(T)$ which is
smoothed on the scale of $\Delta T$, as in Fig.\,\ref{Fig_ChemPot}.
While phenomenological, it successfully accounts for boson
interactions just above $T_c$.  We use the same approach to include
pairon-pairon correlations in the excited states to describe the
fluctuation regime where the correlations persist. The fluctuation
contribution, $S_{fluct}$, can be simply determined by subtracting
the entropy with the standard chemical potential from the entropy
with smoothed chemical potential, Fig.\,\ref{Fig_S fluc}. The
resulting specific heat, Fig.\,\ref{Fig_Cv_num}, shows an almost
identical exponential decay as in the experiments.

\begin{figure}
\includegraphics[width=7.0 cm, trim=10 10 10 10, clip]{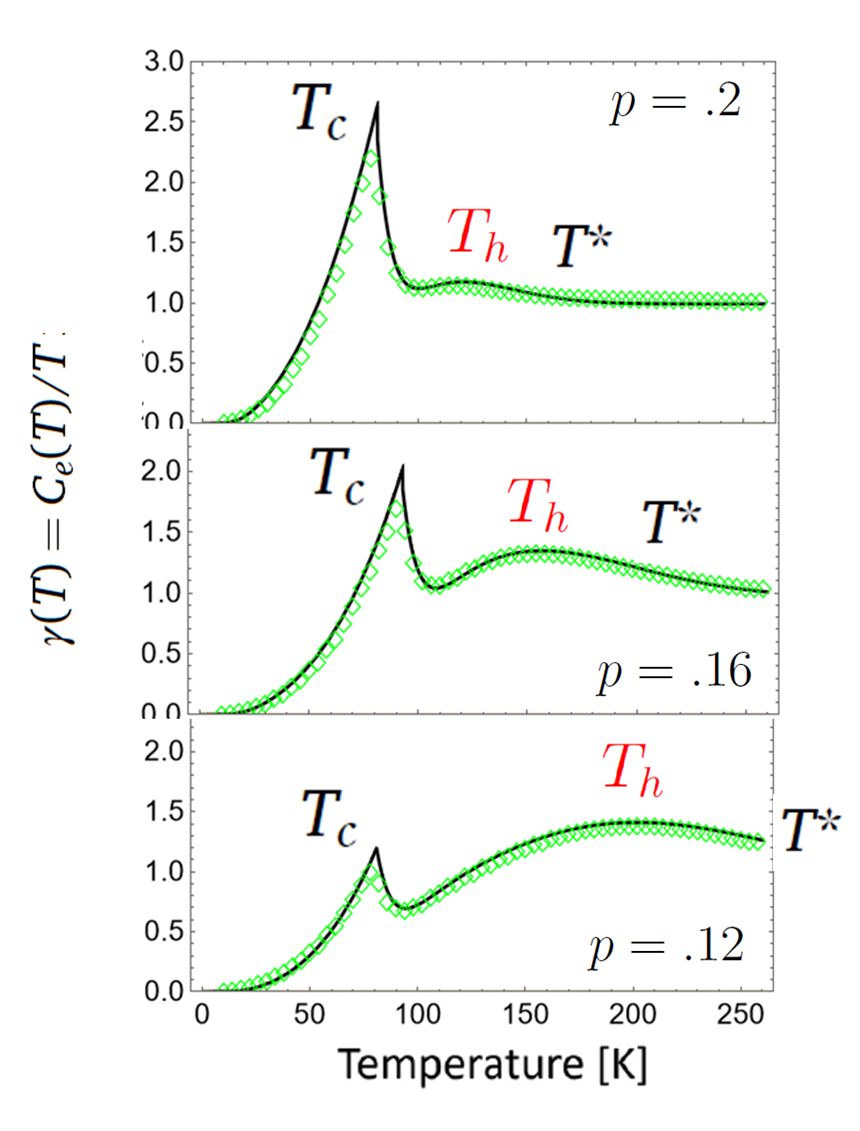}
\includegraphics[width=7.0 cm, trim=10 10 10 10, clip]{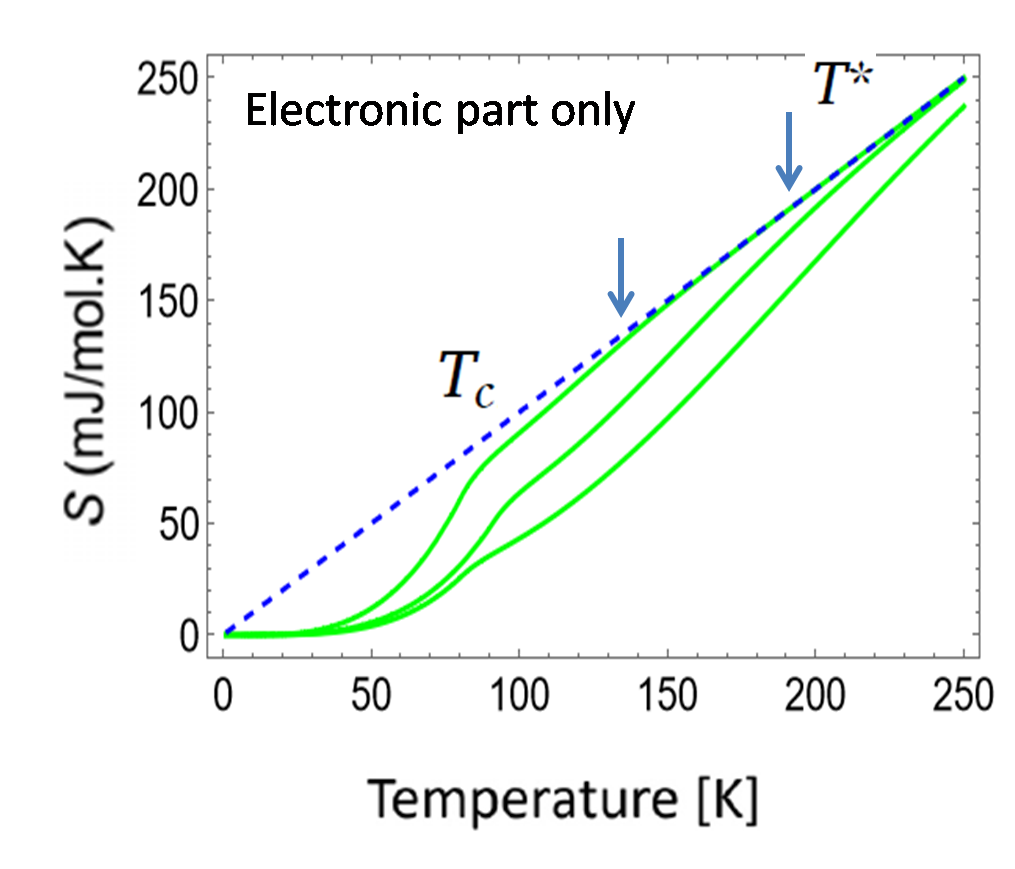}
\caption{(Color online) Temperature dependent entropy and
$\gamma(T)$ coefficient calculated in the framework of the pairon
model \cite{SolStatCom_Noat2021} (values for BSCCO). The modified
chemical potential is used to include the fluctutions (see text).
\\ Upper panel: $\gamma(T)$ coefficient for three different hole
concentrations, underdoped ($p=0.12$), optimally doped ($p=0.16$),
and overdoped ($p=0.2$). The numerical solution (green points) is
compared to the phenomenological expression (black line).
\\ Lower panel: corresponding entropy for the same three hole
concentrations.} \label{Fig_3doping}
\end{figure}

The full calculation of the electronic entropy is done for three
different hole concentrations, $p=0.12$ (underdoped),  $p=0.16$
(optimally doped) and  $p=0.2$ (overdoped). As shown in
Fig.\,\ref{Fig_3doping}, the $\gamma(T)$ coefficient exhibits the
expected peak at $T_c$ due to the vanishing of the condensate. Just
above the peak, the remarkable exponential tail extends from $T_c$
on the characteristic temperature scale $\Delta T\sim 10K$, which we
identify as the fluctuation regime. Beyond $T_c$, $\gamma(T)$ is
still not constant as would be expected in the conventional case.
Indeed, since excited pairs exist above $T_c$, the normal state is
only recovered approaching $T^*$, which is well above $T_c$,
especially in the underdoped case.

We also note that a clear hump is observed below $T^*$ in
$\gamma(T)$ at the characteristic temperature labeled by $T_h$.  As
described in our previous work \cite{SolStatCom_Noat2021}, it
corresponds to the inflection point in the energy gap function
$\Delta_p(T)$. The latter physical parameter, fundamental to the
model, expresses the density of excited pairons which decays with
increasing temperature.

The values of the relevant parameters can be deduced by fitting the
experimental data, which is generally a difficult task. To proceed,
we introduce a simple phenomenological expression for the entropy,
which is the sum of different contributions, the condensate and
excited pairs terms, and the fluctuation term:
\begin{equation}
S_{elec} = S_{cond} + S_{pair} + S_{fluc}
\label{Equa_Selec}
\end{equation}
where
\begin{eqnarray}
&&S_{cond}(T) = S_{pair}(T)\,{\rm e}^{-N_{c}(T)/N_{0}}\ \Theta(T_c-T)\\
&&S_{fluc}(T) = A_{fluc} \, (T-T_c)\,{\rm e}^{-(T-T_c)/\Delta T} \ \Theta(T-T_c) \label{Equa_Sfluc}\\
&&S_{pair}(T) =A_{pair} T\,(1-\alpha_1\,\Delta(T)/\Delta_p)^{1/2} \\
\nonumber
\end{eqnarray}
Where $A_{pair}$ and $A_{fluc}$ are respectively the amplitudes of
the pair term and the fluctuation term, and $\Theta(T)$ is the
standard Heaviside step function. Expression \ref{Equa_Selec} can be
directly compared to the numerical calculation
(Fig.\,\ref{Fig_3doping}, upper panel). Obviously, the agreement is
quite satisfactory.

As we will demonstrate, the fit to the experimental data requires
the contribution of magnetic excitations. Adding the additional
magnetic term, the total entropy is now:
\begin{equation}
S_{tot} = S_{cond} + S_{pair} + S_{fluc} + S_{AF}
\label{Equa_Stot}
\end{equation}
The magnetic entropy can be deduced in a simple way: $S_{AF}
=A_{mag} T\times \chi(T)$, where $A_{mag}$ is the amplitude and
$\chi(T)$ is the magnetic susceptibility of the 2D antiferromagnetic
lattice, calculated in Ref. \cite{JphysChemSol_Lines1970}. The AF
susceptibility can be well described by the simplified
expression\,\cite{Solstatcom_Noat2022}:
\begin{equation}
\chi(T)=\frac{1}{T+\frac{T_{max}^2}{T}+C}
\label{Equa_Chi}
\end{equation}
where $T_{max}$ is the characteristic temperature of magnetic
correlations (where $\chi(T)$ is a maximum) and $C$ is the
Curie-Weiss constant. This simple temperature-dependent expression
was shown to fit successfully the experimental magnetic
susceptibility of cuprates as a function of carrier density
\cite{Solstatcom_Noat2022}.

\section{Analysis of the experimental data}

\subsection{The case of LSCO}
\label{Sec_LSCO} 
\begin{figure}
\includegraphics[width=8.4 cm, trim=10 10 10 10, clip]{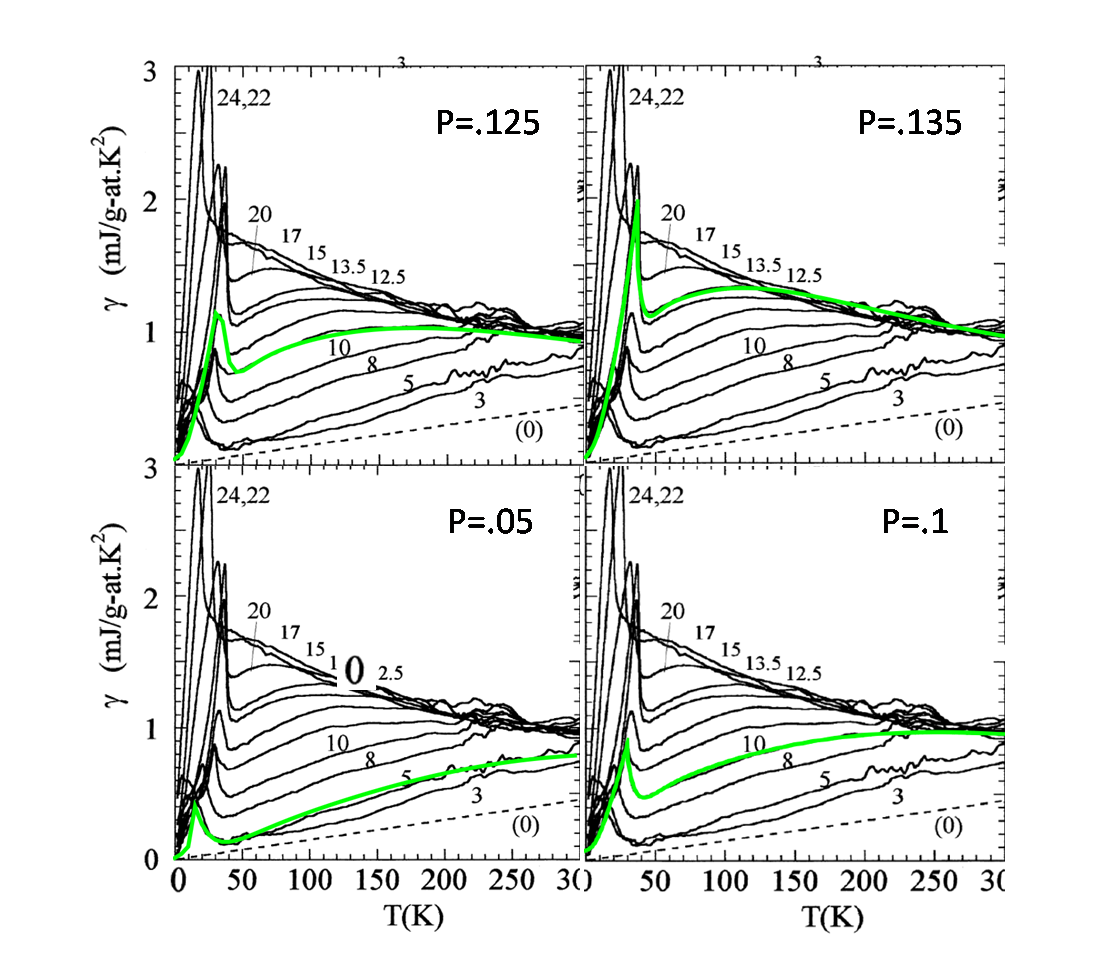}
\caption{(Color online) Experimental $\gamma(T)$ measured in
La$_{2-x}$Sr$_x$CuO$_4$ \cite{JphysChem_Loram2001}. Green lines:
corresponding fits, in the underdoped regime, calculated with the
phenomenological expression. } \label{Fig_Fit_LSCO_Under}
\end{figure}

\begin{figure}
\hskip -.5 cm
\includegraphics[width=9 cm, trim=10 10 10 10, clip]{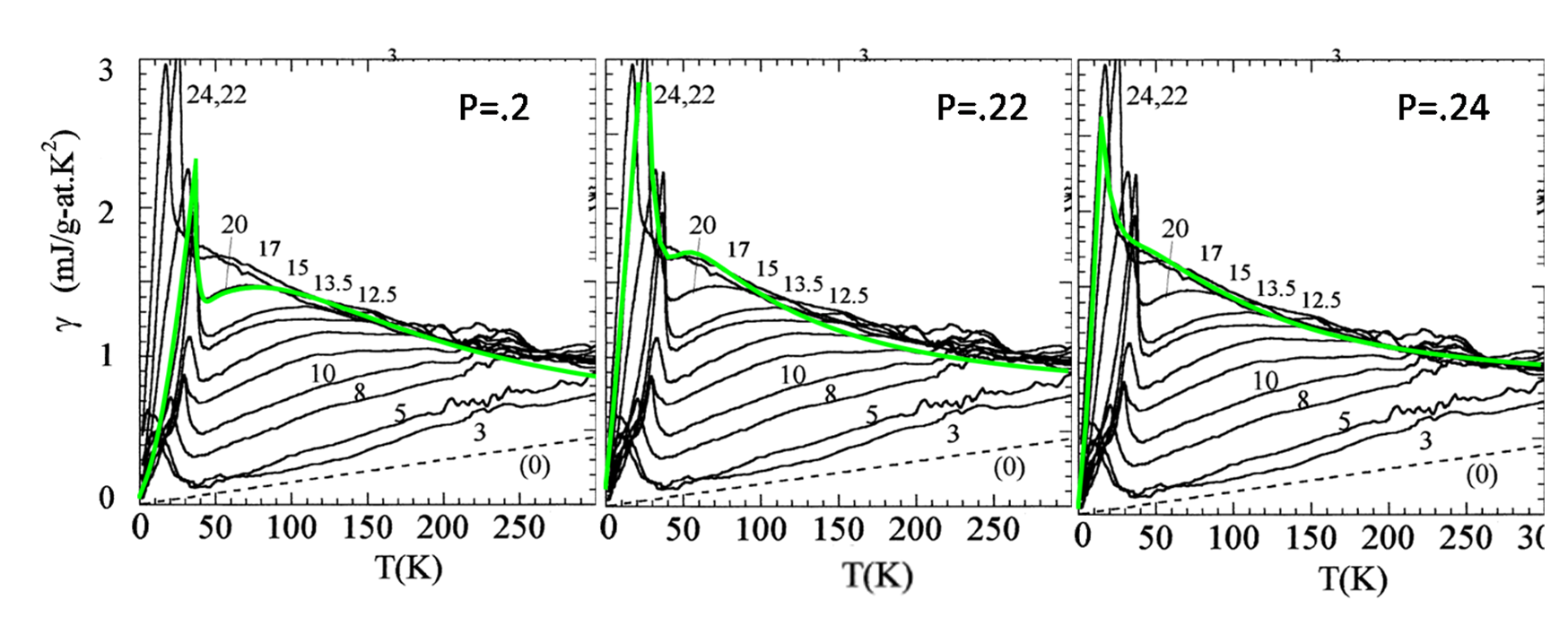}
\caption{(Color online)  Experimental $\gamma(T)$ measured in
La$_{2-x}$Sr$_x$CuO$_4$ \cite{JphysChem_Loram2001} and corresponding
fits in the overdoped regime (green lines). }
\label{Fig_Fit_LSCO_Over}
\end{figure}

We now focus on the fitting and the analysis of
experimental data. We first concentrate on the data obtained by
Loram et al. in LSCO \cite{JphysChem_Loram2001}, see Fig.\,
\ref{Fig_Cv_Exp}. The experimental $\gamma(T)$, and corresponding
fits using the equation \ref{Equa_Stot} for the entropy, are shown
in Fig.\,\ref{Fig_Fit_LSCO_Under} and Fig.\,\ref{Fig_Fit_LSCO_Over}.
The agreement is very satisfactory for the whole doping range.

\begin{figure}[!h]
\includegraphics[width=8.0 cm, trim=10 10 10 10, clip]{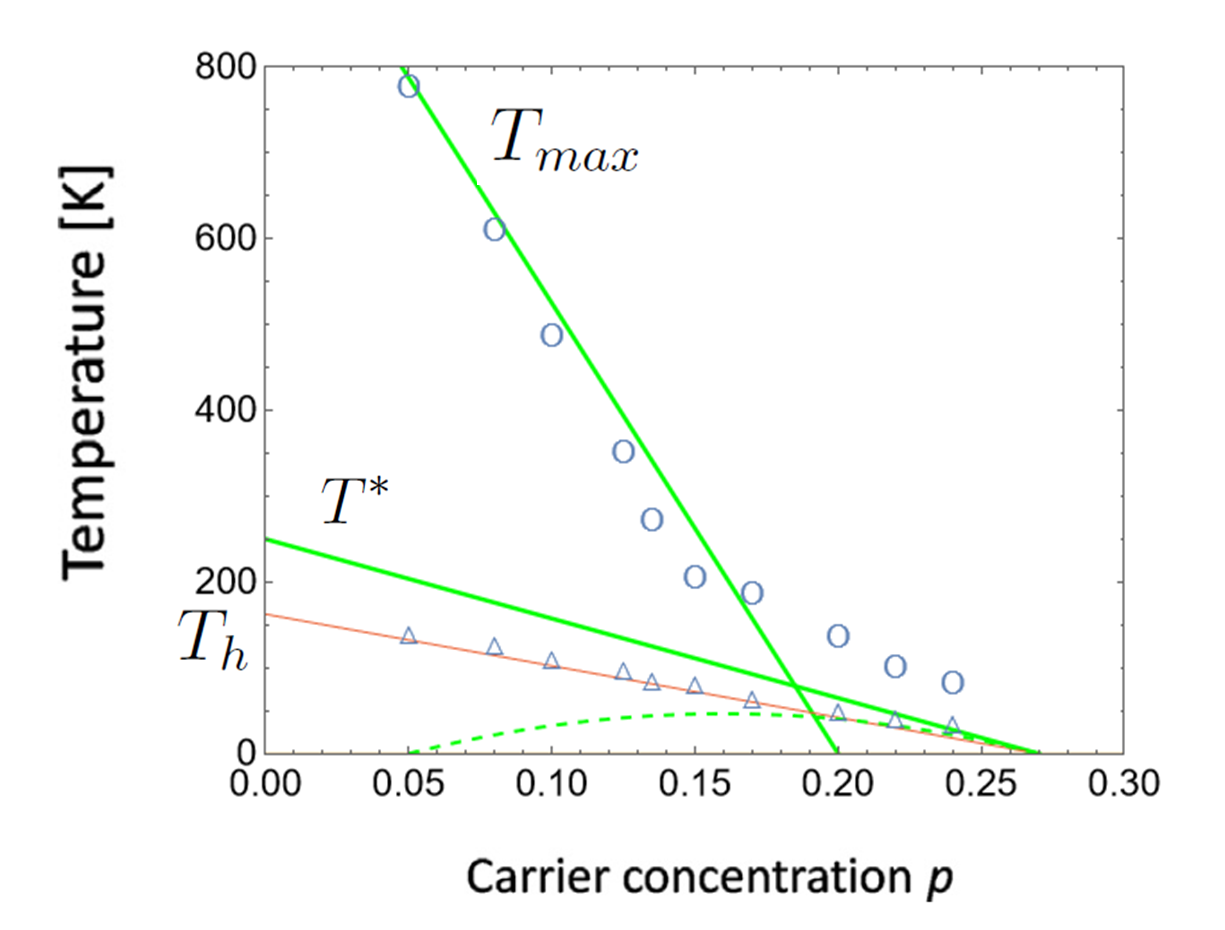}
\caption{(Color online) Phase diagram for La$_{2-x}$Sr$_x$CuO$_4$
deduced from the fits. Above $T_c$, the lines $T_{max}(p)$ and
$T^*(p)$ are clearly distinct. The $T_{max}(p)$ line is in good
agreement with the one deduced from the magnetic susceptibility
\cite{Solstatcom_Noat2022}. From the fits we deduce $T_h(p)$ which
follows $T^*(p)$ with a smaller slope $T_h(p)\approx
\frac{2}{3}T^*(p)$ throughout the doping range. }
\label{Fig_PhaseDiag_LSCO}
\end{figure}

From the fits, we obtain the values of the parameters as a function
of carrier concentration to deduce the phase diagram,
Fig.\,\ref{Fig_PhaseDiag_LSCO}. Above $T_c$, three temperature
scales can be distinguished: the pseudogap temperature $T^*$, the
magnetic temperature $T_{max}$, and the fluctuation temperature
regime, corresponding to the exponential tail observed in
$\gamma(T)$ between $T_c$ and $T_{fluc}=T_c+\Delta T$.

The $T^*$ determined from the fits corresponds roughly to the hump
in the specific heat calculated using the numerical approach: $T_h
\approx (2/3) T^*$. It decreases linearly with carrier concentration
and extrapolates to zero at the end of the $T_c$ dome at $p_{max}$.
From this analysis, there is no indication of the $T^*$ line
crossing the dome.

The magnetic temperature behaves differently. It decreases linearly
in the underdoped regime, clearly seen in Fig.\,
\ref{Fig_PhaseDiag_LSCO}. Remarkably, this linear behavior
extrapolates to zero at $p\sim 0.2$ which is very close to the value
found by Tallon et al. for the vanishing of their gap energy
$E_g(p)$ \cite{JphysChem_Loram2001}. However, the $T_{max}(p)$ curve
derived from the fits deviates from linearity and levels off above
optimum doping ($p \gtrsim .17$), as clearly seen in
Fig.\,\ref{Fig_PhaseDiag_LSCO}. A similar leveling off was deduced
from our analysis of the magnetic susceptibility
\cite{Solstatcom_Noat2022}, in agreement with early pioneering works
\cite{SolstatCom_Oda1990,PRB_Nakano1994}. In short, the analysis of
the specific heat from LSCO confirms the $T_{max}(p)$ magnetic
transition temperature in the phase diagram.

Interestingly, the particular carrier concentration $p\sim 0.2$ also
coincides with a suggested quantum critical point (see
\cite{PRB_girod} and references therein). In our previous work
(\cite{Solstatcom_Noat2022}, \cite{PhysLettA_Noat2022}) we proposed
that the particular slope of the $T_{max}(p)$ line, which
extrapolates down to the same critical value, indicates the
formation of `simplons', or holes surrounded by 4 frozen spins in
the 2D lattice. The symmetry properties of such spin-charge objects
were explored further in our theoretical analysis
\cite{ModelSimul_Noat2022}, in particular their relation to pairon
formation below $T^*$. Whether or not the `simplon-pairon' model is
compatible with some critical transition taking place at this
special carrier density remains an open question.

\subsection{Interpretation of the entropy}

Given these results on the characteristic temperatures, we now
interpret the specific heat and the entropy in a different light
with respect to Tallon et al.
\cite{JphysChem_Loram2001,FrontPhys_Tallon2022}, who have focused on
YBCO (Fig.\,\ref{Fig_Cv_Exp}, panel (a)), and to some extent on
BSCCO (panel (c)).

\begin{figure}
\includegraphics[width=9 cm, trim=10 10 10 10, clip]{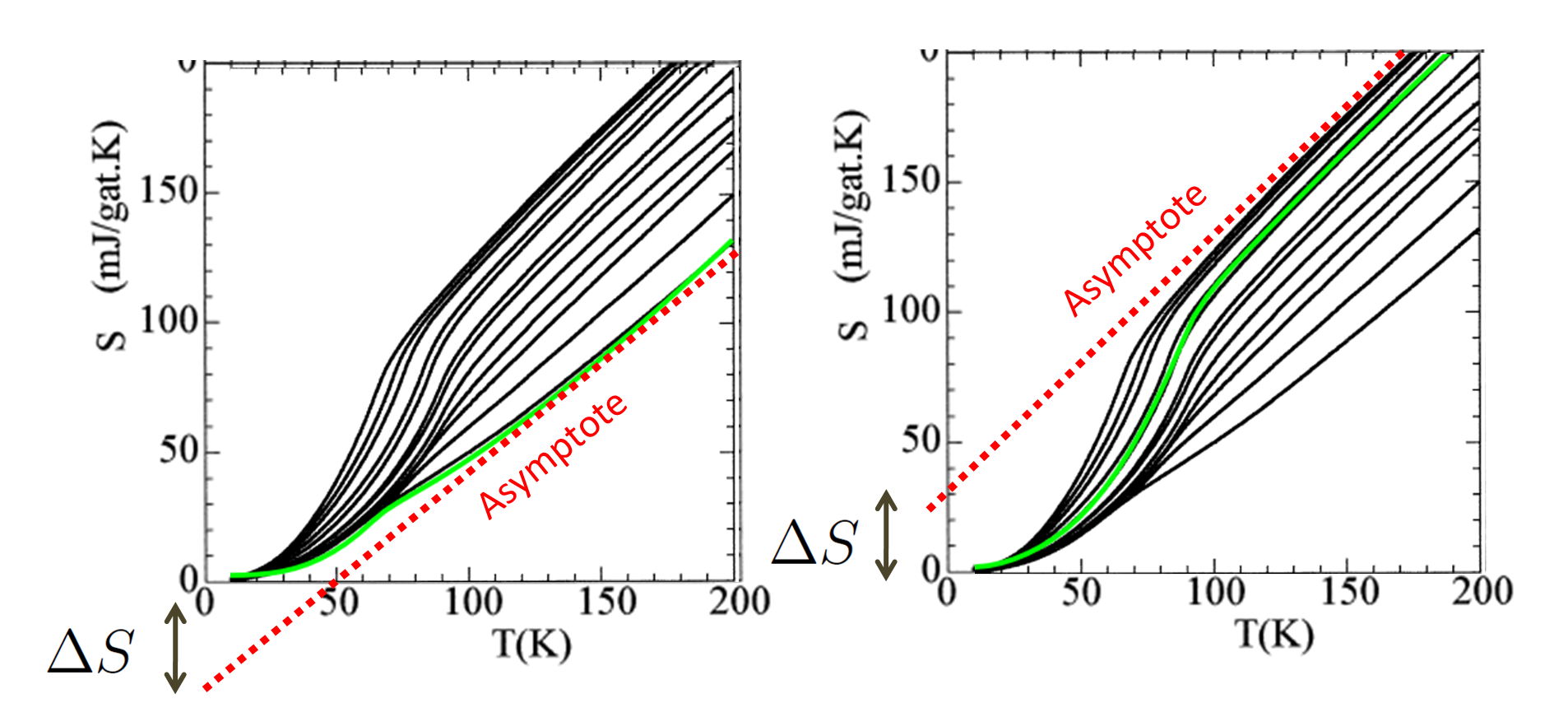}
\caption{(Color online) Left panel: entropy of
Bi$_2$Sr$_2$CaCu$_2$O$_{8+\delta}$ in the underdoped regime (adapted
from Ref. \cite{JphysChem_Loram2001}). Right panel:  entropy in the
overdoped regime and corresponding fits (green lines). In the
underdoped regime the asymptotic line (red dash line) is below the
normal state $\gamma(T)$, while it is above in the overdoped regime
(adapted from Ref. \cite{JphysChem_Loram2001}). This entropy shift
is due to the contribution of magnetic excitations and not to a gap
in the electronic DOS, as described in the text.} \label{Fig_BSCCO}
\end{figure}

As mentioned previously, these authors consider the `pseudogap' as
being due to a constant electronic gap in the Fermi level DOS, at
fixed $p$, of magnitude $E_g(p)$. In this model, one does not
recover the entropy of the normal state $S(T) \sim \gamma\, T$ for
large $T \gg T_c$. Instead, the entropy line is shifted to a lower
value $S(T) \sim \gamma T - \Delta S$, as seen in
Fig.\,\ref{Fig_BSCCO}, left panel for BSCCO. They deduce the gap
energy from $E_g(p) \propto \Delta S/k_B$ and plot the corresponding
phase diagram as a function of $p$. A similar $E_g(p)$ was also
deduced from the magnetic susceptibility of LSCO and YBCO
\cite{PhysicaC_Naqib2007,ScSciTech_Naqib2008,JScNovMag_Islam2010}
which, according to their model, vanishes near $p_c\simeq 0.2$.

\begin{figure*}
\centering \vbox to 6cm{
\includegraphics[width=14 cm, trim=10 10 10 10, clip]{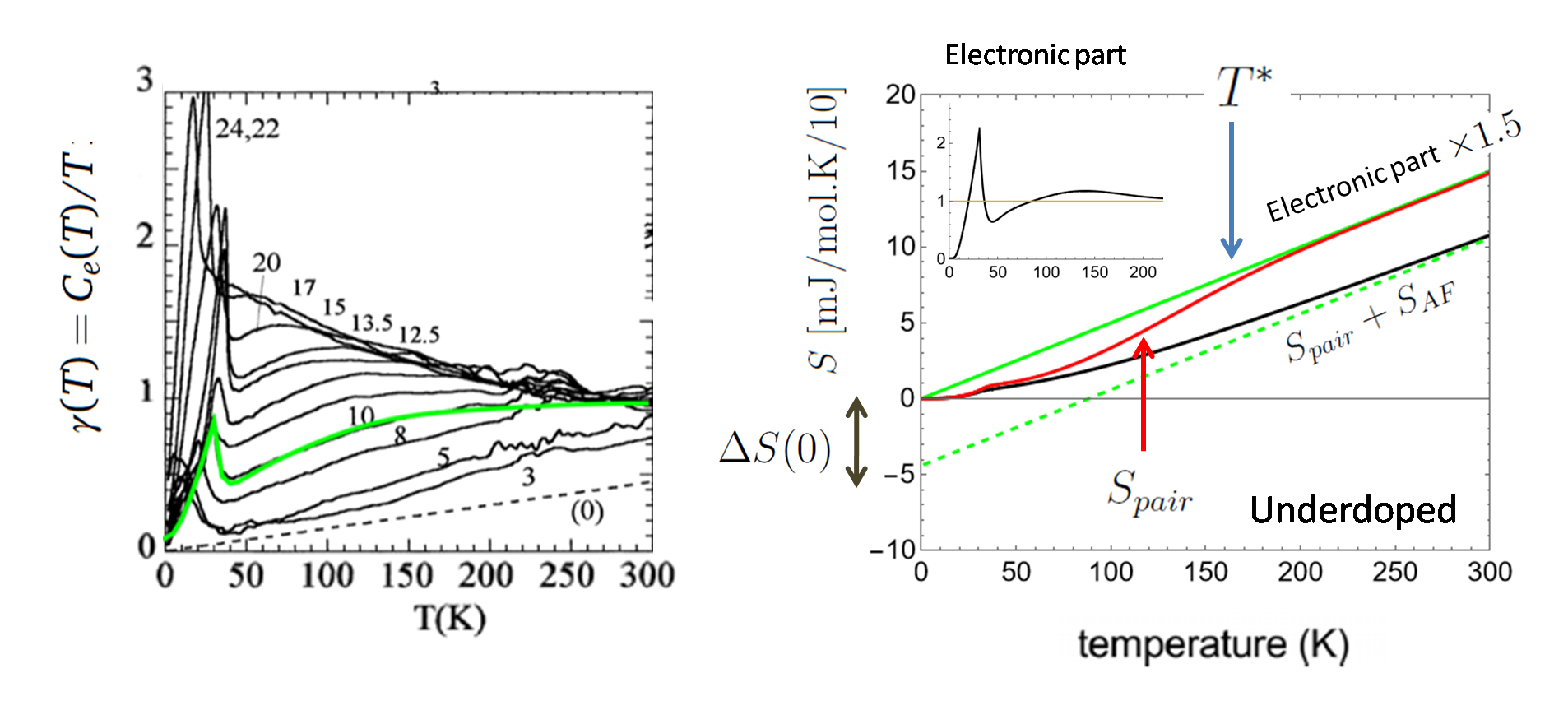}
} \caption{(Color online) Left panel: experimental $\gamma(T)$ in
the underdoped regime ($p=0.1$) and corresponding fit (green line).
Right panel: corresponding total entropy calculated from the fit
(black line) and electronic part (red line). The magnetic term
causes an apparent shift in the entropy. The apparent asymptotic
line is below the normal $\gamma_N \times T$ entropy. Inset:
corresponding electronic $\gamma(T)$ obtained by substracting the
magnetic part. } \label{Fig_magneticPart_under}
\end{figure*}
\begin{figure*}
\centering \vbox to 6cm{
\includegraphics[width=14 cm, trim=10 10 10 10, clip]{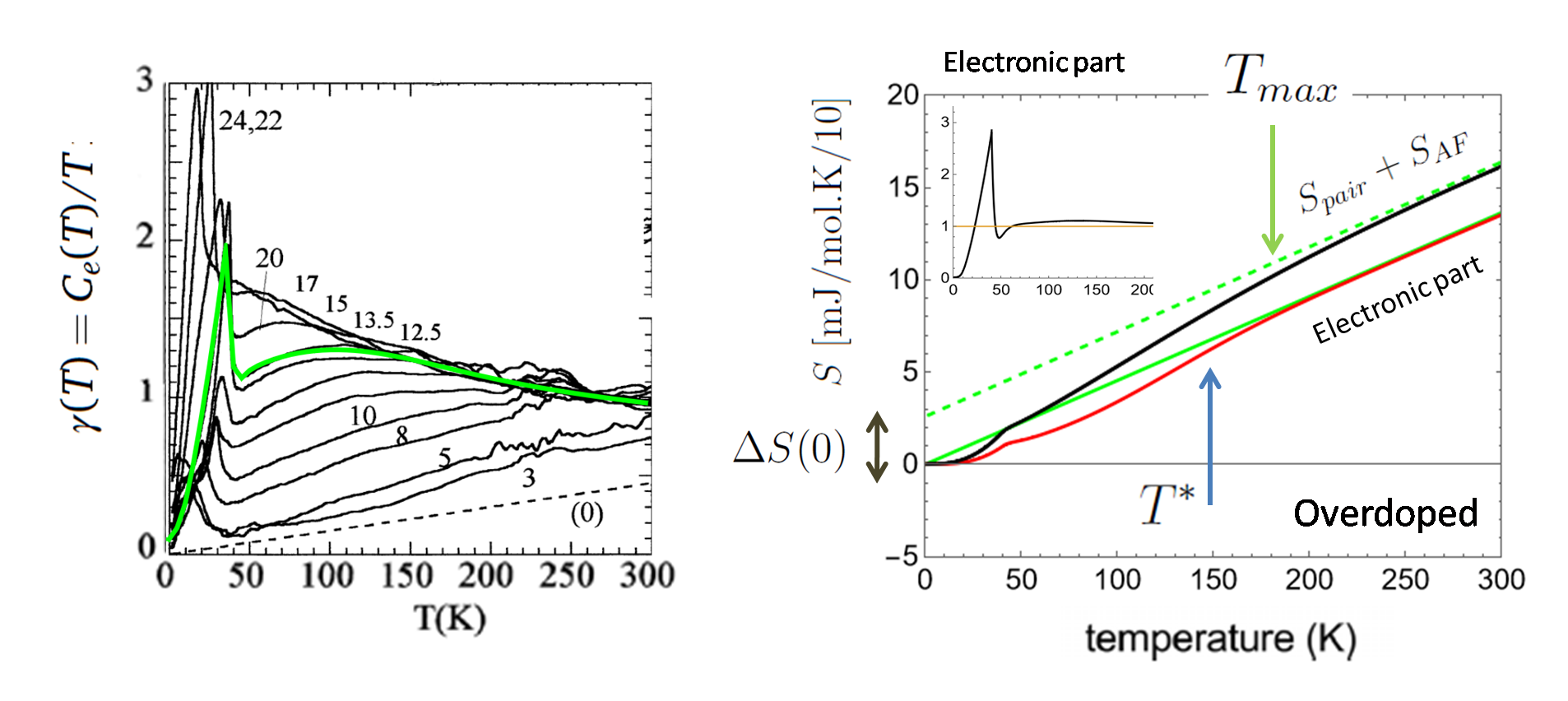}
} \caption{(Color online) Left panel: experimental $\gamma(T)$ in
the slightly overdoped regime (hole doping $p=0.17$) and
corresponding fit (green line). Left panel: corresponding entropy
calculated from the fit (black line) and electronic part (red line).
The apparent asymptotic line is above the normal $\gamma_N\times T$
entropy. Inset: corresponding electronic $\gamma(T)$ obtained by
substracting the magnetic part. } \label{Fig_magneticPart_over}
\end{figure*}

Several difficulties emerge from this interpretation: firstly, the
gap $E_g$ does not close with rising {\it temperature} and secondly,
the characteristic line $E_g(p)/k_B$ crosses the dome in the phase
diagram. In addition, and perhaps most significantly, it fails to
explain the behavior of $S(T)$ in the overdoped regime where $\Delta
S$ becomes positive, as in Fig.\,\ref{Fig_BSCCO}, right panel.
Indeed, a positive $\Delta S$ cannot be interpreted in terms of a
`gap' since $E_g$ would have the wrong sign.

This question can be further explored thanks to the general formula
Eq.\ref{Equa_Stot} for the entropy. Indeed, it confirms that the
entropy shift, $\Delta S$, is due to the magnetic part leading to
the correct evolution of $S(T)$ with carrier density. In the case of
BSCCO this conclusion is clearly illustrated in
Fig.\,\ref{Fig_BSCCO} where we have fitted the entropy from Loram et
al. using the {\it identical equation} (\ref{Equa_Stot}) as in the
previous case of LSCO. For two contrasting hole concentrations,
respectively in the underdoped and overdoped regimes, we see that
the dashed red line illustrates well the asymptotes of the $S(T)$
curves. In the underdoped case, this line extrapolates to a negative
$\Delta S$ at $T=0$, while in the overdoped case, it extrapolates to
a positive value. Clearly this entropy shift effect is continuous as
a function of carrier density.

The general formula for $S(T)$ conveniently allows to separate the
magnetic and electronic contributions to the total entropy. What's
more, it allows to follow the $S(T)$ profile while varying the key
parameters ($A_{pair}, A_{mag}, T^*, T_{max}$, etc.) around their
`best fit' values. This parameter tweaking can equally be done on
the fits for Fig.\,\ref{Fig_BSCCO}, as well as the previous series
of fits for LSCO, Figs.\,\ref{Fig_Fit_LSCO_Under} and
\ref{Fig_Fit_LSCO_Over}. Exploring the parameter space in this way
gives a novel insight on the influence of the AF magnetic term on
the overall shape of the entropy above $T_c$.

In the pairon model the strictly electronic part of the entropy must
return to the normal state, i.e. $\sim \gamma_N\,T$, once pairing
vanishes at $T^*$. To confirm this important property, we return to
the case of LSCO to extract the entropy terms in $S(T)$ using the
equation (\ref{Equa_Stot}). The electronic and magnetic terms of the
entropy can be isolated from the best fits: underdoped case
Fig.\,\ref{Fig_magneticPart_under} and overdoped case
Fig.\,\ref{Fig_magneticPart_over}. In the right panel of both
figures, the black line corresponds to the total entropy, while the
red line corresponds to the isolated electronic part.

Consider first the underdoped case,
Fig.\,\ref{Fig_magneticPart_under}. The asymptotic dotted green line
extrapolates to a negative value at $T=0$, resulting in an apparent
negative shift $\Delta S$, as previously noted for BSCCO. Once the
magnetic part is removed, the electronic part, red line, follows the
expected behavior: the normal state is recovered above $T^*$ due to
pair breaking. To the contrary, in the overdoped case,
Fig.\,\ref{Fig_magneticPart_over}, the asymptotic dotted line
extrapolates to a positive $\Delta S$ value at $T=0$. Again, once
the magnetic part is removed, red curve, the normal state entropy is
recovered above $T^*$.

To summarize, the change of the $S(T)$ asymptote from underdoped to
overdoped is due to the magnetic $T\chi(T) $ term in the entropy,
which depends strongly on the characteristic temperature $T_{max}$.
In the underdoped regime, for $T<T_{max}$, the magnetic part gives
rise to an additional contribution to the entropy $S \sim \gamma
T-\Delta S$ in this temperature range. On the other hand, in the
overdoped regime, the apparent entropy shift changes sign. The
latter is thus not due to a constant ($T$-independent gap) in the
electronic DOS, but to the contribution of magnetic excitations to
the entropy. Moreover, for any carrier concentration, our results
show that the strictly {\it electronic} part of the total $S(T)$
reveals the $T$-dependent pairing gap and has the normal $\sim
\gamma_N\,T$ asymptote above $T^*$ as required.

\begin{figure}[!h]
\includegraphics[width=8 cm, trim=10 10 10 10, clip]{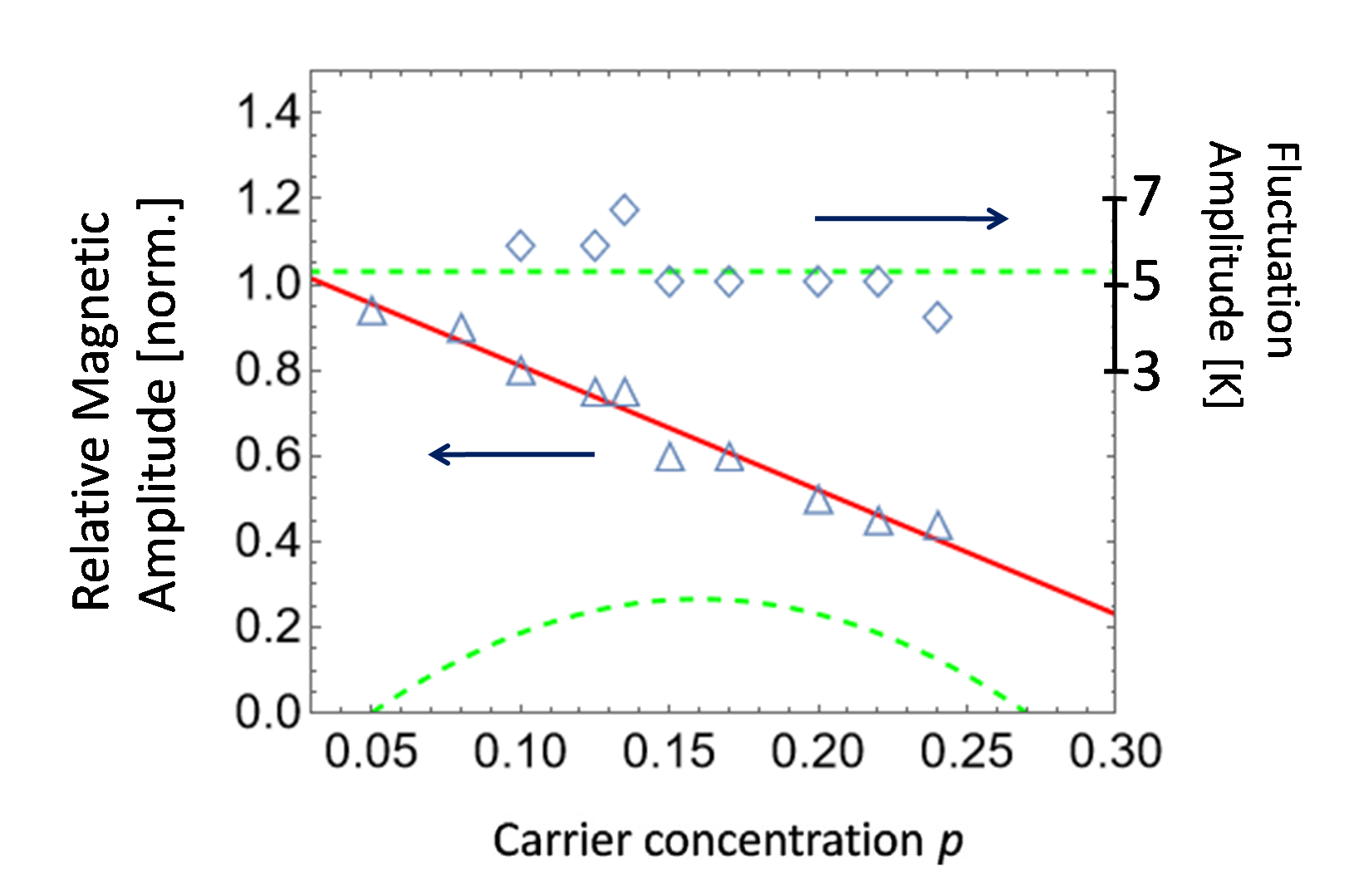}
\caption{(Color online) Relative magnetic amplitude
$\frac{A_{mag}}{A_{mag}+A_{pair}}$ deduced from the fits of
$\gamma(T)$ in La$_{2-x}$Sr$_x$CuO$_4$ and fluctuation temperature
scale $\Delta T$ as a function of hole concentration $p$.}
\label{Fig_Amp_LSCO}
\end{figure}

From the fits, we also find a novel linear dependence of the
magnetic relative amplitude $A_{mag}/(A_{mag}+A_{pair})$ as a
function of carrier concentration, which decreases from underdoped
to overdoped sides of the phase diagram (see
Fig.\,\ref{Fig_Amp_LSCO}). However, the detailed shape of the
entropy is still very sensitive to the parameters, in particular
both the $T_{max}$ and $T^*$ values, which are much lower in the
overdoped case, as shown in the phase diagram
Fig.\,\ref{Fig_PhaseDiag_LSCO}. This explains why the `hump' in the
background in $\gamma(T)$ is much closer to the critical
temperature, and may be difficult to resolve without further
analysis. These remarks help explain why the correct antinodal
pseudogap is difficult to pinpoint in the specific heat, and indeed
many other thermal/transport measurements, as compared to ARPES and
tunneling. We thus give a clear response to the objections of Tallon
et al. \cite{PRB_Tallon2023} who argue that, in the underdoped case,
the entropy never recovers the normal state even at high temperature
due to the pseudogap, contradicting the `pairon' model.

The magnetic entropy of antiferromagnetic material has been given
much attention \cite{JphysChemSol_Lines1970,AdvPhys_deJongh1974}. As
discussed above, our results confirm that it changes significantly
depending on the $T_{max}$ value, which separates spin blocking to
Curie-Weiss fluctuations. In the underdoped regime $T_{max}$ is
large (compared to $T_c$ and $T^*$) and the effect of spin
excitations on $\gamma(T)$ is such that it increases with
temperature (strong AF correlations). At the opposite end of the
dome, in the overdoped regime, $T_{max}$ is much smaller and a
Curie-Weiss law is responsible for the decrease of $\gamma(T)$ with
temperature. These conclusions, independent of the pairon model, are
in agreement with the overall background observed in $\gamma(T)$.

\subsection{The case of YBCO}

Our analysis for the two materials LSCO, single layer, and BSCCO,
double layer, is consistent. The question now arises for the case of
oxygen doped YBCO.

\begin{figure}
\includegraphics[width=8. cm, trim=10 10 10 10, clip]{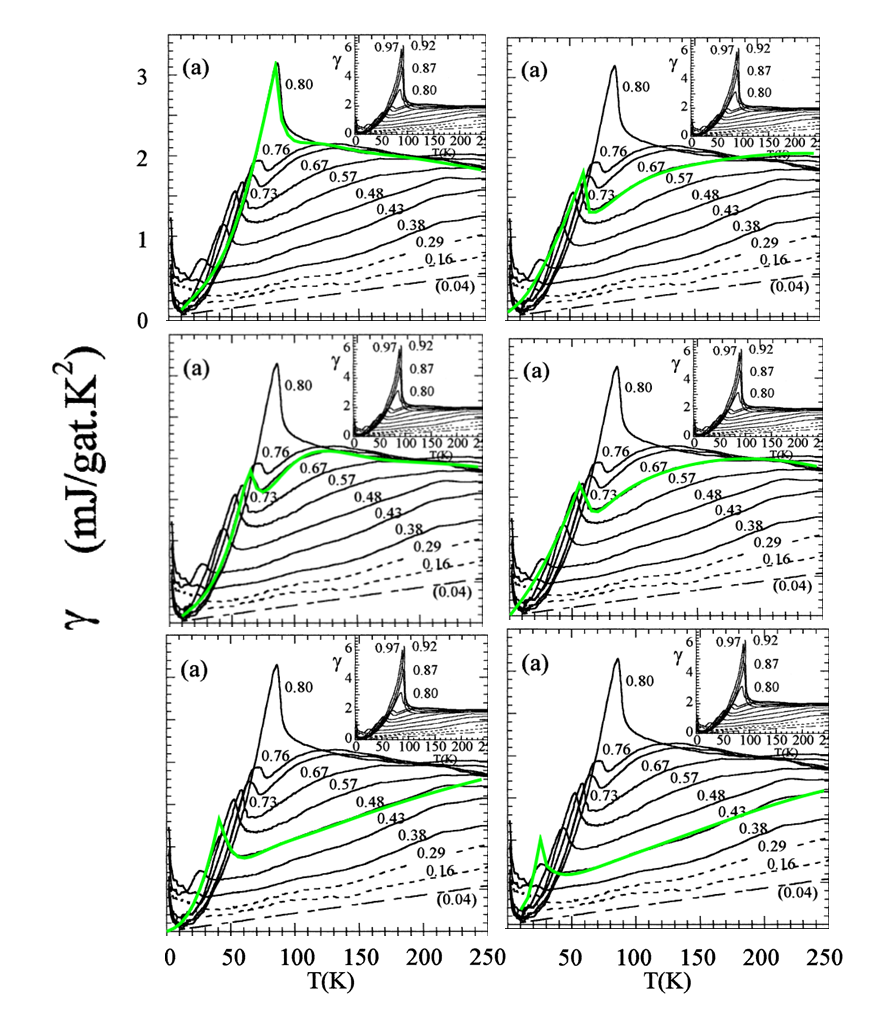}
\caption{(Color online) Experimental $\gamma(T)$ measured in
YBa$_2$Cu$_3$O$_{7-\delta}$ (adapted from Ref.
\cite{JphysChem_Loram2001}) and corresponding fit calculated with
the same phenomenological expression (green lines). }
\label{Fig_Fits_YBCO}
\end{figure}

\begin{figure}
\vbox to 6 cm{
\includegraphics[width=7 cm, trim=10 10 10 10, clip]{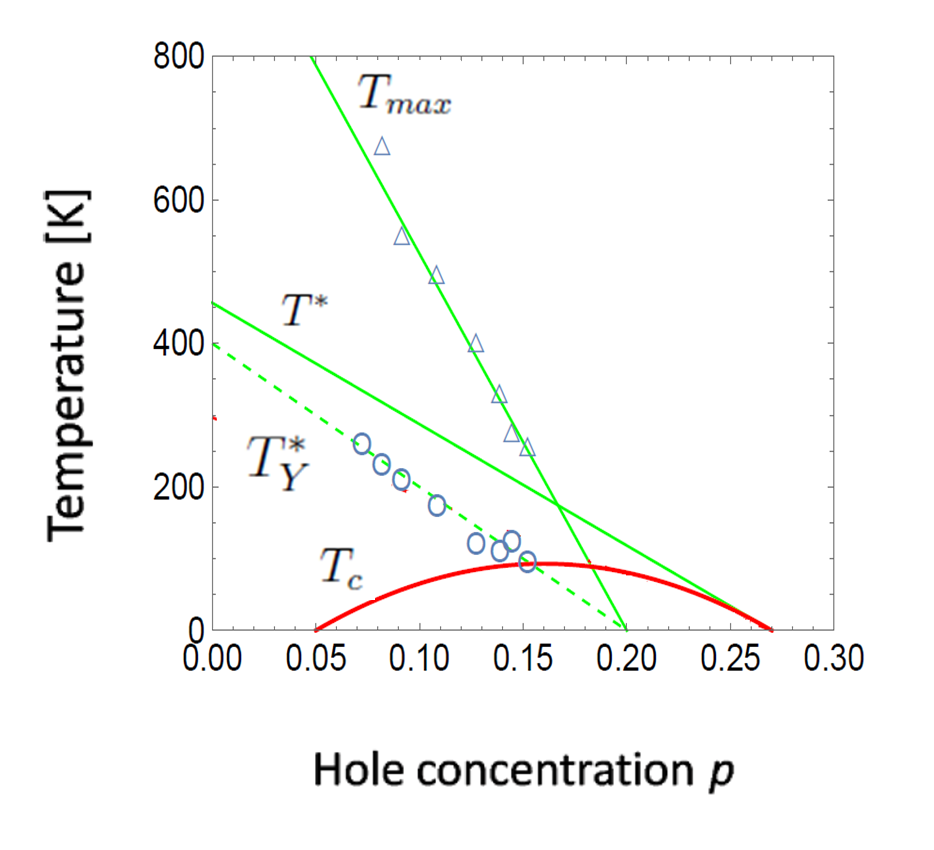}
} \caption{(Color online) Phase diagram of
YBa$_2$Cu$_3$O$_{7-\delta}$ deduced from the fits. Both lines
$T_Y^*(p)$ and  $T_{max}(p)$ seem to converge towards the critical
value $p_c=0.2$. Contrary to the two other materials, the $T_Y^*(p)$
line has a different behavior, as discussed in the text.}
\label{Fig_PhaseDiag_YBCO}
\end{figure}

We then perform the same fitting procedure for oxygen doped YBCO,
Fig.\,\ref{Fig_Fits_YBCO}. As for LSCO, three temperature scales can
be clearly distinguished above $T_c$: the pseudogap temperature
$T_Y^*$, the magnetic temperature $T_{max}$ and the fluctuation
temperature  $T_{fluc}$. Surprisingly, we find that the phase
diagram, shown in Fig.\,\ref{Fig_PhaseDiag_YBCO}, turns out to be
different from the one obtained for LSCO. The two temperatures
$T_Y^*(p)$ and $T_{max}(p)$, while remaining distinct, seem to
converge towards the same point $p_c\approx 0.2.$ and eventually
vanish there.

The apparent pseudogap temperature deduced from the fits follows the
line:
\begin{equation}
2.2\, k_BT_Y^*(p)\simeq 2.2\,k_BT^*(p)-A\times \left(\frac{p}{p_c}
\right) \label{Eq_modTstar}
\end{equation}
where $A=21.5$ is a constant, $p_c\simeq 0.2$ and $T^*(p)$ is the
expected pseudogap temperature line (found for BSCCO). The results
of the fits clearly show that YBCO is different from the two other
materials, although they mostly have the same relevant physical
parameters. As is well known, YBCO is different in its structure
(see \cite{FrontPhys_Hott2004} for a review); it is more isotropic
than LSCO and BSCCO and has a combination of CuO planes and 1D
chains \cite{APL_Beno1987}. In addition to the specific heat, both
the magnetic susceptibility and resistivity show a different
behavior, which remains controversial. The progressive filling of
the oxygen chains parallel to the 2D CuO planes with oxygen doping
may explain the unusual $T_Y^*(p)$ line. For example, this
progressive filling could preferentially modify the Fermi-level gap
along the nodal direction, hence leading to the modified law
Eq.\ref{Eq_modTstar}. Further studies are necessary to resolve this
apparent paradox.

\subsection{Unconventional fluctuation regime}

As noted previously, the specific heat appears to have a common
signature of an unusually large exponential decay above the critical
temperature. From the fits in LSCO and YBCO, we have extracted the
characteristic temperature scale $\Delta T$ of the fluctuation
regime above $T_c$. It is important to note that these fluctuations
do not commence below $T_c$ as in the model of Tallon et
al.\cite{PRB_Tallon2011}. This effect is clearly illustrated in the
present model in Figs.\,\ref{Fig_ChemPot},\ref{Fig_S
fluc},\ref{Fig_Cv_num},\ref{Fig_3doping}.

\begin{figure}[!h]
\includegraphics[width=8. cm, trim=10 10 10 10, clip]{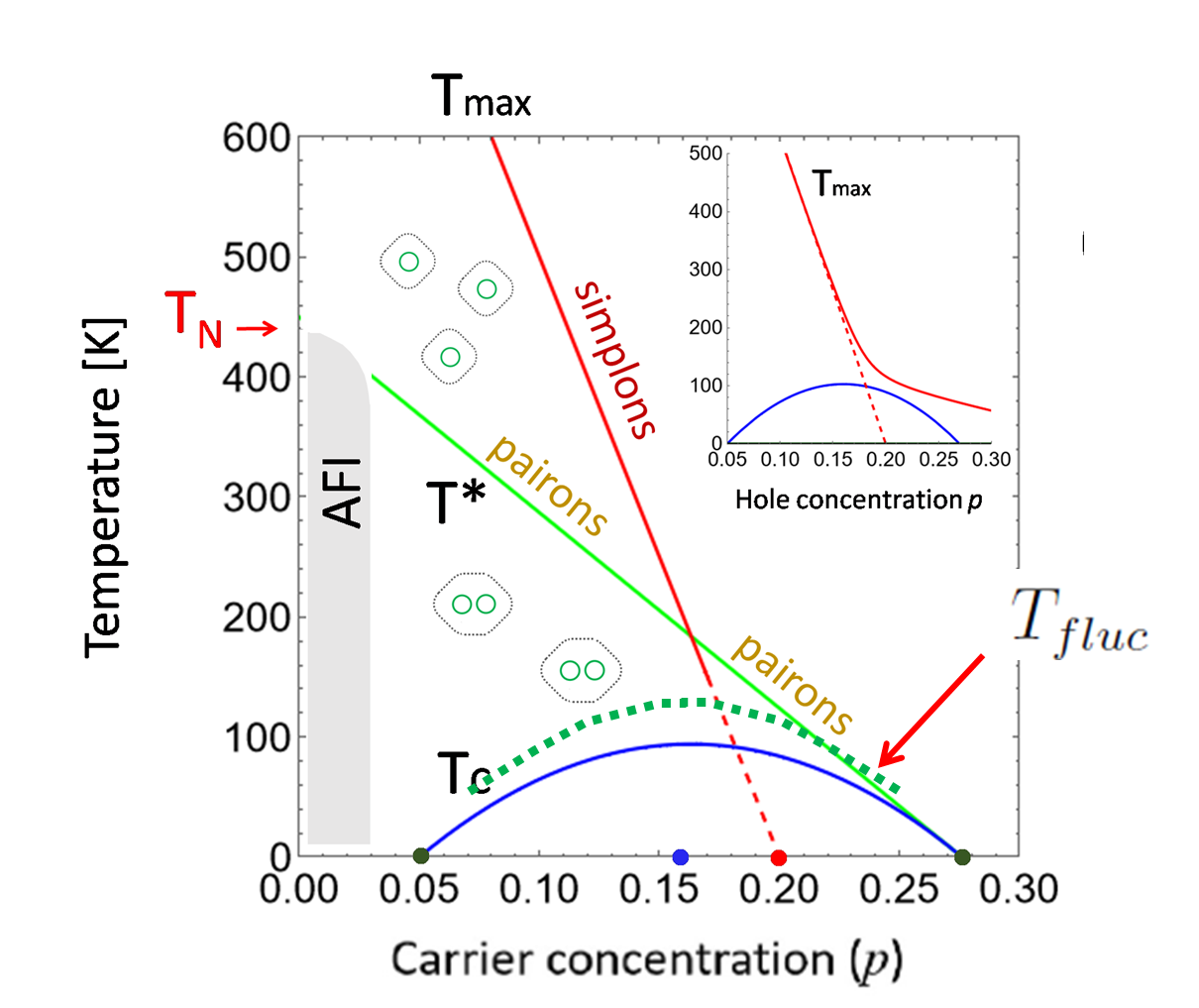}
\caption{(Color online) Generic phase diagram of
La$_{2-x}$Sr$_x$CuO$_4$ and Bi$_2$Sr$_2$CaCu$_2$O$_{8+\delta}$ (the
numerical values for $T_c$ and $T^*$ are the for
Bi$_2$Sr$_2$CaCu$_2$O$_{8+\delta}$; the $T_{max}(p)$ line is for
both materials). $T^*(p)$ is the characteristic temperature of
magnetic correlations. Below $T_{max}$, `simplons', one hole
surrounded by 4 frozen spins are formed. Pairons form at the onset
temperature $T^*$, the pseudogap temperature. They condensed below
the critical temperature $T_c$. Above $T_c$, we indicate the
unconventional fluctuation regime (as discussed in the text) on the
scale $\Delta T$, which follows the critical dome. }
\label{Fig_Generic}
\end{figure}

In view of the difference in $T_c$, it can be noted that $\Delta
T\sim 5K$ for LSCO and twice that value for YBCO and BSCCO (see
Fig.\,\ref{Fig_Amp_LSCO} for the case of LSCO). The scale of the
fluctuation regime is therefore much larger than the BCS case (a few
percent of $T_c$), see Fig.\,\ref{Fig_Cv_Val}. Furthermore, $\Delta
T$ hardly varies with doping and therefore $T_c+\Delta T$ follows
the dome, as in Fig.\,\ref{Fig_Generic}. This finding is in
qualitative agreement with the result of Tallon et al., although
obtained with a different analysis \cite{PRB_Tallon2011}.

Since the fluctuation regime is clearly unconventional, we suggest a
novel mechanism. In the pairon model, the minimum hole density to
have a condensate $p_{min}\simeq 0.05$, at the beginning of the
$T_c$-dome, corresponds to one pairon every $\sim 6-7 a_0$ where
$a_0$ is the lattice constant. This is related in our model to the
fundamental pairon-pairon interacting distance $d_0$, responsible
for the condensation.

These considerations imply a new energy scale in the phase diagram:
\begin{equation}
2 \delta_M \approx J_{eff} \times p_{min}
\end{equation}
where  $J_{eff}$ is the effective antiferromagnetic exchange energy
at $p=p_{min}$, Numerically, it gives the approximate value $2
\delta_M\approx 3.8$  meV, clearly smaller than any previously
mentioned energy scale.

On the other hand, as in the derivation of the entropy, it is
necessary to have a minimum activation of a pairon from the
condensate $\delta\sim 1-2$ meV due to the Bose-Einstein singularity
in a 2D system \cite{SciTech_Sacks2015}. A remarkable coincidence is
that the mini-gap in the excitation spectrum is the same order of
magnitude as the fluctuation energy $\delta_M$ described above.
Hypothetically, these unconventional fluctuations are due to the
activation of pairons in and out of the condensate across the
mini-gap once the density has reached a critically small value. This
is clearly in support of a new and important energy scale in
cuprates, $\delta_M$.

\section{Conclusion}

In this work, we revisit the specific heat of cuprates in the
framework of the pairon model. In particular, we introduce a general
expression for the temperature-dependent entropy. While no
contradiction with the pairon model is found, our analysis strongly
suggests that it is crucial to take into account the magnetic
contribution.

The general shape of the cuprates entropy can be well described by
the model. From precise fits for three different cuprates, LSCO,
BSCCO and YBCO, we have deduced the three temperature scales are
present above $T_c$ in the phase diagram: the pseudogap temperature
$T^*$, the magnetic temperature $T_{max}$, and the fluctuation
temperature $T_{fluct}$. The resulting phase diagram is completely
consistent with the one previously deduced from the magnetic
susceptibility.

We show that a strong evolution of the entropy from the underdoped
to the overdoped regime is mainly due to the contribution of the
magnetic excitations and not to a $T$-independent gap in the
electronic density of states. However, a temperature dependent
pseudogap, consistent with a pairing gap, does exist throughout the
phase diagram. Its temperature scale $T^*(p)$ for LSCO and BSCCO
follows a linear behavior and vanishes and the end of the dome, in
agreement with photoemission and tunneling spectroscopy. A different
behavior is found for YBCO, where the apparent pseudogap temperature
follows a steeper linear law and vanishes near $p_c\simeq 0.2$,
similar to $T_{max}(p)$.

The fluctuation regime is found to be highly unconventional, the
characteristic temperature scale of fluctuations being large and
almost independent of carrier density. We relate this effect to the
minimum doping value $p_{min}$, at the onset the $T_c$-dome. It is
consistent with the mini-gap $\delta_M \sim 1-2meV$ in the pairon
excitation energy spectrum that is required for the condensation
mechanism. It is therefore a key energy scale of the problem.

\vskip 2 mm

\section{Acknowledgements}

The authors gratefully acknowledge discussions with Dr. Hiroshi
Eisaki, Dr. Shigeyuki Ishida, Dr. Jeff Tallon, and Prof. Atsushi
Fujimori.


\vskip 2 mm

\begin{thebibliography}{99}

\bibitem{ChiPhysB_Wen2020}  Hai-Hu Wen , Specific heat in superconductors, Chinese Phys. B {\bf 29} 017401 (2020).

\bibitem{PR_BCS1957} J. Bardeen, L. Cooper, J. Schrieffer, Theory of Superconductivity, Phys. Rev. {\bf 108} 1175 (1957).

\bibitem{JPhysCondMatt_Klimczuk2012} T. Klimczuk, M. Szlawska, D. Kaczorowski, J. R. O'Brien and D. J. Safarik, Superconductivity in the Einstein solid
VAl$_{10.1}$, J. Phys.: Condens. Matter {\bf 24}, 365701 (2012).

\bibitem{PRB_Moca2002} C. P. Moca and Boldizs\'ar Jank\'o, Electronic specific heat in the pairing pseudogap regime, Phys. Rev. B  {\bf 65}, 052503 (2002).

\bibitem{PRL_Curty2002} Philippe Curty and Hans Beck, Thermodynamics and Phase Diagram of High Temperature Superconductors, Phys. Rev. Lett. {\bf 91}, 257002 (2003).

\bibitem{PRB_Borne2010} A. J. H. Borne, J. P. Carbotte, and E. J. Nicol, Specific heat across the superconducting dome in the cuprates, Phys. Rev. {\bf B 82}, 094523 (2010).

\bibitem{PRL_Loram1993} J. W. Loram, K. A. Mirza, J. R. Cooper, and W. Y. Liang, Electronic specific heat of YBa$_2$Cu$_3$O$_{6+x}$ from 1.8 to 300 K, Phys. Rev. Lett.  {\bf 71}, 1740 (1993).

\bibitem{PhysicaC_Loram1994} J.W.Loram, K.A.Mirza, J.M.Wade, J.R.Cooper, W.Y.Liang, The electronic specific heat of cuprate superconductors, Physica C  {\bf 235--240}, Pages 134 (1994).

\bibitem{JphysJapSol_Matzusaki2004} Toshiaki Matsuzaki, Naoki Momono, Migaku Oda and Masayuki Ido, Electronic Specific Heat of La$_{2-x}$Sr$_x$CuO$_4$: Pseudogap Formation and Reduction of the Superconducting Condensation Energy,
Journal of the Physical Society of Japan, {\bf 73}, 2232 (2004).

\bibitem{PRL_Wen2009} Hai-Hu Wen, Gang Mu, Huiqian Luo, Huan Yang, Lei Shan, Cong Ren, Peng Cheng, Jing Yan, and Lei Fang,
Specific-Heat Measurement of a Residual Superconducting State in the Normal State of Underdoped Bi$_2$Sr$_{2-x}$La$_x$CuO$_{6+\delta}$ Cuprate Superconductors,
Phys. Rev. Lett. {\bf 103}, 067002 (2009).

\bibitem{PRB_Inderhees1987} S. E. Inderhees, M. B. Salamon, T. A. Friedmann, and D. M. Ginsberg, Measurement of the specific-heat anomaly at the superconducting transition of YBa$_2$Cu$_3$O$_{7-\delta}$ Phys. Rev. B {\bf 36}, 2401(R) (1987).

\bibitem{SolStatCom_Noat2021} Y. Noat, A. Mauger, M. Nohara, H. Eisaki, W. Sacks
, How `pairons' are revealed in the electronic specific heat of cuprates, Solid State Communications {\bf 323}, 114109 (2021).

\bibitem{PRB_Inderhees1988} S. E. Inderhees, M. B. Salamon, Nigel Goldenfeld, J. P. Rice, B. G. Pazol, D. M. Ginsberg, J. Z. Liu, and G. W. Crabtree,
Specific heat of single crystals of YBa$_2$Cu$_3$O$_{7-\delta}$: Fluctuation effects in a bulk superconductor
Phys. Rev. Lett. {\bf 60}, 1178 (1988);

\bibitem{JPhysChemSol_Loram1998} J.W.Loram, K.A.Mirza, J.R.Cooper, J.L.Tallon, Specific heat evidence on the normal state pseudogap, Journal of Physics and Chemistry of Solids {\bf 59}, 2091(1998).

\bibitem{JphysChem_Loram2001} J.W. Loram, J. Luo, J.R. Cooper, W.Y. Liang, J.L. Tallon, Evidence on the pseudogap and condensate from the electronic specific heat, Journal of Physics and Chemistry of Solids {\bf 62}, 59 (2001).

\bibitem{FrontPhys_Tallon2022} Jeffery L. Tallon, James G. Storey, Thermodynamics of the pseudogap
in cuprates, Front. Phys. 10:1030616 (2022).

\bibitem{Revmod_Fisher2007} \O. Fischer, M. Kugler, I. Maggio-Aprile,
C. Berthod and C. Renner, Scanning tunneling spectroscopy of the
cuprates, Rev. Mod. Phys. {\bf 79}, 353 (2007).

\bibitem{Nat_Hashimoto2014} Makoto Hashimoto, Inna M. Vishik, Rui-Hua He, Thomas P. Devereaux and Zhi-Xun Shen, Energy gaps in high-transition-temperature cuprate superconductors, Nature Physics {\bf 10}, 483 (2014).

\bibitem{LowTemp_Kordyuk2015} A. A. Kordyuk, Pseudogap from ARPES experiment:Three gaps in cuprates and topological superconductivity, Low Temp. Phys. {\bf 41}, 319 (2015).

\bibitem{RepProgPhys_Hufner2008} S. H\"ufner, M. A. Hossain, A Damascelli, and G. A. Sawatzky,Two gaps make a high-temperature superconductor?, Rep. Prog. Phys., {\bf 71}, 062501 (2008).

\bibitem{Solstatcom_Noat2022}  Y. Noat, A. Mauger, M. Nohara, H. Eisaki, W. Sacks, Cuprates phase diagram deduced from magnetic susceptibility: what is the 'true' pseudogap line?, Solid State Communications {\bf 348--349}, 114689 (2022).

\bibitem{PRB_Torrance1989} J. B. Torrance, A. Bezinge, A. I. Nazzal, T. C. Huang, S. S. P. Parkin, D. T. Keane, S. J. LaPlaca, P. M. Horn, and G. A. Held, Properties that change as superconductivity disappears at high-doping concentrations in La$_{2-x}$Sr$_x$CuO$_4$, Phys. Rev. B {\bf 40}, 8872 (1989).

\bibitem{SolstatCom_Oda1990} M.Oda, H.Matsuki, M.Ido, Common features of magnetic and superconducting properties in Y-doped Bi$_2$(Sr,Ca)$_3$Cu$_2$O$_8$ and Ba(Sr)-doped La$_2$CuO$_4$, Solid State Communications {\bf 74}, 1321 (1990).

\bibitem{PRB_Nakano1994} T. Nakano, M. Oda, C. Manabe, N. Momono, Y. Miura, and M. Ido, Magnetic properties and electronic conduction of superconducting  La$_{2-x}$Sr$_x$CuO$_4$, Phys. Rev. B {\bf 49}, 16000 (1994).

\bibitem{PRB_Tallon2023} Jeffrey L. Tallon and James G. Storey, Thermodynamics and the pairon model for cuprates, Phys. Rev. B {\bf 107}, 054507 (2023).

\bibitem{EPL_Sacks2017} W. Sacks, A. Mauger and Y. Noat, Cooper pairs without glue in high-$T_c$ superconductors: A universal phase diagram, Euro. Phys. Lett {\bf 119}, 17001 (2017).

\bibitem{SciTech_Sacks2015}  W. Sacks, A. Mauger, Y. Noat, Pair\,--\,pair interactions as a mechanism for
high-T$_c$ superconductivity, Superconduct. Sci. Technol., {\bf 28}
105014 (2015).

\bibitem{PhysLettA_Noat2022} Yves Noat, Alain Mauger, William Sacks. Superconductivity in cuprates governed by topological constraints. Physics Letters A {\bf 444}, 128227 (2022).

\bibitem{ModelSimul_Noat2022} Yves Noat  Alain Mauger, William Sacks, Statistics of the cuprate pairon states on a square lattice, Modelling Simul. Mater. Sci. Eng. {\bf 31}, 075010 (2023).

\bibitem{EPJB_Sacks2016} William Sacks, Alain Mauger, and Yves Noat, Unconventional temperature dependence of the cuprate excitation spectrum, Eur. Phys. J. B {\bf 89}, 183 (2016).

\bibitem{Jphys_Sacks2018} William Sacks, A. Mauger and Y. Noat, Origin of the Fermi arcs in cuprates: a dual role of quasiparticle and pair excitations, Journal of Physics: Condensed Matter, {\bf 30},  475703 (2018).

\bibitem{EPL_Noat2019} Yves Noat, Alain Mauger and William Sacks, Single origin of the nodal and antinodal gaps in cuprates,
Euro. Phys. Lett {\bf 126}, 67001 (2019).

\bibitem{Adv_Kocharovsky2006} Vitaly V. Kocharovsky, Vladimir V. Kocharovsky, Martin Holthaus, C.H. Raymond Ooi, Anatoly Svidzinsky, Wolfgang Ketterle, Marlan O. Scully, Fluctuations in Ideal and Interacting Bose-Einstein Condensates: From the Laser Phase Transition Analogy to Squeezed States and Bogoliubov Quasiparticles, Advances In Atomic, Molecular, and Optical Physics {\bf 53}, 291-411 (2006).

\bibitem{JphysChemSol_Lines1970} M. E. Lines, The quadratic-layer antiferromagnet,
J. Phys. Chem. Solids . {\bf 31}, 101 (1970).

\bibitem{PRB_girod}C. Girod, D. LeBoeuf, A. Demuer, G. Seyfarth, S. Imajo, K. Kindo, Y. Kohama,
M. Lizaire, A. Legros, A. Gourgout, H. Takagi, T. Kurosawa, M. Oda,
N. Momono, J. Chang, S. Ono, G.-q. Zheng, C. Marcenat, L. Taillefer,
and T. Klein, Normal state specific heat in the cuprate
superconductors La$_{2-x}$Sr$_{x}$CuO$_{4}$ and
Bi$_{2+y}$Sr$_{2-x-y}$La$_{x}$CuO$_{6+\delta}$ near the critical
point of the pseudogap phase, Phys. Rev. B {\bf 103}, 214506 (2021),
and Michon, B., Girod, C., Badoux, S. et al., Thermodynamic
signatures of quantum criticality in cuprate superconductors, Nature
{\bf 567}, 218 (2019).

\bibitem{PhysicaC_Naqib2007} S.H.Naqib, J.R.Cooper, Effect of the pseudogap on the uniform magnetic susceptibility of Y$_{1?x}$Ca$_x$Ba$_2$Cu$_3$O$_{7?\delta}$, Physica C: Superconductivity {\bf 460--462}, 750-752 (2007).

\bibitem{ScSciTech_Naqib2008} S H Naqib and R S Islam, Extraction of the pseudogap energy scale from the static magnetic susceptibility of single and double CuO$_2$ plane high-T$_c$ cuprates,Supercond. Sci. Technol. {\bf 21}, 105017 (2008).

\bibitem{JScNovMag_Islam2010} R.S. Islam, M.M. Hasan, S.H. Naqib, Nature of the Pseudogap in High-T c Cuprates: Analysis of the Bulk Magnetic Susceptibility of La$_{2-x}$ Sr$_x$Cu$_{1-y}$Zn$_y$O$_4$, J. Supercond. Nov. Magn.  {\bf 23}, 1569 (2010).

\bibitem{AdvPhys_deJongh1974} L.J. de Jongh and A. R. Miedema, Experiments on simple magnetic model systems, Adv. Phys. {\bf 23}, 1 (1974).

\bibitem{FrontPhys_Hott2004} Roland Hott, Reinhold Kleiner, Thomas Wolf and Gertrud Zwicknag, Superconducting Materials -- A Topical Overview, in Frontiers in Superconducting Materials, Ed. Anant V. Narlikar, Springer Verlag, Berlin, pp 1-69 (2004).

\bibitem{APL_Beno1987} M. A. Beno, L. Soderholm, D. W. Capone, D. G. Hinks, J. D. Jorgensen, J. D. Grace; Ivan K. Schuller, C. U. Segre; K. Zhang, Structure of the single phase high temperature superconductor YBa$_2$Cu$_3$O$_{7-\delta}$, Appl. Phys. Lett.  {\bf  51}, 57 (1987).

\bibitem{PRB_Tallon2011} J. L. Tallon, J. G. Storey, and J. W. Loram, Fluctuations and critical temperature reduction in cuprate superconductors,
Phys. Rev. B  {\bf 83}, 092502 (2011).

\end{thebibliography}
\end{document}